\documentstyle[sprocl]{article}

\input{psfig}

\def\V{{\cal V}}
\def\L{{\cal L}}
\def\D{{\cal D}}

\def\dalpha{{\dot{\alpha}}}
\def\dbeta{{\dot{\beta}}}
\def\dmu{{\dot{\mu}}}
\def\dnu{{\dot{\nu}}}
\def\bbeta{\bar{\eta}}
\def\bxi{\bar{\xi}}
\def\bchi{\bar{\chi}}
\def\btheta{\bar{\theta}}
\def\bphi{\bar{\phi}}
\def\bsigma{\bar{\sigma}}
\def\blambda{\bar{\lambda}}
\def\bQ{\bar{Q}}
\def\bD{\bar{D}}
\def\bU{\bar{U}}
\def\bE{\bar{E}}

\def\be{\begin{equation}}
\def\ee{\end{equation}}
\def\bea{\begin{eqnarray}}
\def\eea{\end{eqnarray}}

\catcode`\@=11 
\newcommand{\lsim}{\mathrel{\mathpalette\@versim<}}
\newcommand{\gsim}{\mathrel{\mathpalette\@versim>}}
\def\@versim#1#2{\vcenter{\offinterlineskip
 \ialign{$\m@th#1\hfil##\hfil$\crcr#2\crcr\sim\crcr } }}
 \def\cropen#1{\crcr\noalign{\vskip #1}}
\def\crr{\cropen{1\jot }}
\catcode`\@=12 

\newcounter{mysection}
\newcounter{mysectionstar}
\newcommand{\mysection}[1]{\setcounter{mysectionstar}{0}
\stepcounter{mysection}\setcounter{equation}{0}
\par\bigskip\noindent{\bf \themysection.\ \ #1}\nopagebreak[4]\par\vskip .3cm}
 
\begin{document}

\title{WEAK-SCALE SUPERSYMMETRY:\\[2mm]
THEORY AND PRACTICE}

\author{JONATHAN A. BAGGER}

\address{Department of Physics and Astronomy\\
Johns Hopkins University\\
Baltimore, MD 21218}

\maketitle\abstracts{
These lectures contain an introduction to the theory and practice
of weak-scale supersymmetry.  They begin with a discussion of the
hierarchy problem and the motivation for weak-scale supersymmetry.
They continue by developing the coset approach to superfields.  They
use superfield techniques to construct the minimal supersymmetric
version of the standard model and to discuss soft supersymmetry
breaking and its implications.  The lectures end with a brief
survey of expectations for future collider experiments.}

\mysection{Introduction and Motivation}

During the past decade, the standard model of particle physics
has been tested to a remarkable degree of accuracy.  Precision
measurements have confirmed its predictions to the level of
radiative corrections \cite{langacker} -- \cite{lep working group}.  
With the discovery of the top quark, the matter sector of the standard
model stands essentially complete.  All that remains is to find the
Higgs, the missing ingredient of the standard model.

Such is the conventional wisdom.  In reality, the situation
is not so simple.  While there is no doubt that precision tests have
challenged the standard model as never before, the status of the Higgs
is still open to question.  At present, the experimental limits do
not reveal much about the Higgs and its properties \cite{pdg}.  
Indeed, many theorists believe that the search for the Higgs will
uncover new physics that is even more interesting than that associated
with the Higgs itself.

These beliefs are motivated by a host of theoretical problems with
the ordinary standard model.  Perhaps the most compelling is the
so-called hierarchy problem, the famous instability of the Higgs
mass under quadratically divergent radiative corrections
\cite{hierarchy}.  These lectures will explain the hierarchy problem
and use it to motivate a new symmetry -- called supersymmetry -- that
might become manifest at the TeV scale \cite{motivation} -- \cite{mybook}.
If supersymmetry is correct, it will lead to a rich new spectroscopy
in the years to come.

It is in this spirit that these lectures will present an introduction
to weak-scale supersymmetry.  (They will not discuss physics at the
Planck scale.)  They will develop the necessary supersymmetric technology
and use it to construct the minimal supersymmetrized version of the
standard model.  They will also prepare the ground for the lectures
of Tata \cite{tata} and Seiberg \cite{seiberg}.

We shall start by discussing the hierarchy problem.  To understand
the issues involved, we will consider a toy model with one complex
scalar, $A$, and one Weyl fermion, $\chi$.  We take the Lagrangian
to be as follows,
\begin{eqnarray}
\L \; = & - & \partial_m A^* \partial^m A\;-\;i \bchi \bsigma^m
\partial_m \chi \nonumber \\
& - & {1\over2}\,M_F\, \chi\chi\;-\;{1\over2}\,M_F\, \bchi\bchi \;- 
\;\lambda_F\, A\chi\chi\;-\;\lambda_F \,A^* \bchi\bchi \nonumber \\[1mm]
& - & M^2_B\, A^*A\;-\; \lambda_B\, (A^*A)^2\;,
\label{toy model}
\end{eqnarray}
where we use two component spinor notation, outlined in the
Appendix.

\begin{figure}[t]
\hspace*{1.1truein}
\psfig{figure=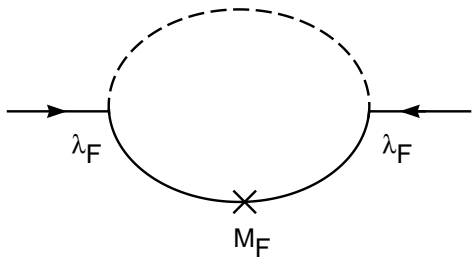,height=1.5in}
\caption{The one-loop correction to the fermion mass is logarithmically
divergent, and proportional to the fermion mass, $M_F$.}
\label{Fmass}
\end{figure}

The Lagrangian (\ref{toy model}) enjoys a global U(1) chiral
symmetry,
\begin{eqnarray}
A & \rightarrow & e^{-2 i \alpha}\, A \nonumber \\
\chi & \rightarrow & e^{i \alpha}\, \chi \;.
\end{eqnarray}
This symmetry is broken only by the fermion mass, $M_F$.  Because
of this symmetry, the one-loop fermion mass correction must contain
at least one mass insertion, as shown in Fig.~\ref{Fmass}.
Therefore the fermion mass correction is multiplicative, of
the form
\begin{equation}
\delta M_F\;\simeq\;{\lambda_F^2\over 16 \pi^2}\;M_F\;.
\label{delta mf}
\end{equation}
Equation (\ref{delta mf}) illustrates why fermion masses are said
to be {\it natural:}  they are stable under radiative corrections.  Once 
$M_F$ is fixed at tree level, it is protected from large radiative
corrections by the U(1) chiral symmetry \cite{hierarchy}.

\begin{figure}[t]
\hspace*{0.8truein}
\psfig{figure=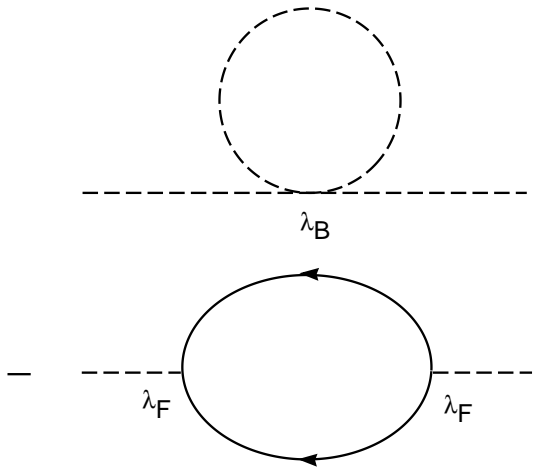,height=2.5in}
\caption{The one-loop corrections to the boson mass contain two
quadratically divergent contributions of opposite sign.}
\label{Bmass}
\end{figure}

The boson mass, $M_B$, stands in contrast to the fermion mass.
The boson mass is not protected by the chiral symmetry, so at one loop,
it receives additive contributions, as shown in Fig.~\ref{Bmass}.   
By power counting, one finds that the scalar mass renormalizations
are quadratically divergent,
\begin{equation}
\delta M_B^2\; =\; {\lambda_B\over16 \pi^2}\, \Lambda^2\;
-\;{\lambda_F^2 \over16 \pi^2}\, \Lambda^2 \;,
\label{boson one loop}
\end{equation}
where $\Lambda$ is a large ultraviolet cutoff, and the minus sign
comes from the fermion loop. Equation (\ref{boson one loop})
illustrates why light scalar masses are {\it not natural.}  Their
tree-level values are not stable; they receive large, quadratically
divergent, radiative corrections \cite{hierarchy}.  

For the case of the standard model, this analysis applies to the
scalar Higgs boson, $h$.  In the standard model, the Higgs mass,
$M_h$, is of order the $W$ mass, $M_W$, and is
proportional to a vacuum expectation value, $v$.  The
vev $v^2$ receives quadratically divergent radiative corrections.  
This means that the natural scale for the Higgs mass is of order
the cutoff, $\Lambda$, which is presumably the Planck scale, $M_P$,
or the unification scale, $M_{\rm GUT}$.

Of course, technically speaking, there is nothing wrong with
this instability.  It is certainly possible to  adjust the one-loop
counterterms so that they cancel the quadratic divergence.  However,
this cancellation requires an exquisite fine tuning of one part in
$10^{17}$ to maintain the hierarchy $M_W \ll M_P$.  This fine
tuning is not natural; it lies at the heart of the hierarchy problem.

The toy model discussed above illustrates the hierarchy problem, but
it also hints at a possible resolution.  From eq.~(\ref{boson one loop})
we see that it is possible for the quadratic divergences to cancel between
the bosonic and fermionic loops.  For the case at hand, this requires
that $\lambda_B$ be related to $\lambda_F^2$.  More generally -- and 
to ensure that the cancellation persists to all orders -- it requires
a symmetry, called supersymmetry.

During the course of these lectures, we shall see that supersymmetry
protects the hierarchy $M_W \ll M_P$ by canceling all 
dangerous quadratic divergences.
In the supersymmetric standard model, this requires a doubling of
the particle spectrum.  For every particle that has been discovered,
supersymmetry predicts another that has not.  The extra particles
circulate in loops and protect the hierarchy from destabilizing
divergences \cite{motivation}.

In what follows we will also review present expectations for the
supersymmetric particle spectrum.  We will see that current limits
pose no serious constraints on the parameter space.  
We will also see that the next generation of accelerators, including
the Fermilab Main Injector, LEP 200, and a possible higher-luminosity
Tevatron, will open a new era in supersymmetric particle searches.
These accelerators will -- for the first time -- begin to probe
significant regions of the supersymmetric parameter space.  And with
the advent of the LHC, we shall find that weak-scale supersymmetry
will be placed to a definitive test.

\mysection{Supersymmetry and the Wess-Zumino Model}

Supersymmetric field theories are based on the following algebra
\cite{mybook},
\begin{eqnarray}
\{ Q_\alpha, \;\bQ_\dalpha\} & = & 2
\sigma_{\alpha \dalpha}^m P_m \nonumber\\
\{ Q_\alpha, \; Q_\beta \} & = & 
\{ \bQ_\dalpha, \;\bQ_\dbeta\} \; = \; 0 \nonumber \\
\left[ P_m, \; Q_\alpha \right] & = & \left[ P_m, \; \bQ_\dalpha
\right] \; = \; 0 \nonumber \\
\left[ P_m,\; P_n \right] & = & 0\;.
\label{susy alg}
\end{eqnarray}
This is a graded Lie algebra because it contains bosonic and fermionic
generators.  (In four dimensions, there can be up to eight fermionic
generators $Q^A_\alpha$, with $A = 1,...,8.$  We shall restrict our
attention to the simplest case, with only one generator, $Q_\alpha$.)

The supersymmetry algebra relates particles of different spins.  It
is a nontrivial extension of the usual Poincar\'e spacetime symmetry.  
Indeed, the local version of supersymmetry leads to an extension of
Einstein gravity, called supergravity \cite{sugra}.  Supergravitational
effects are suppressed by powers of $M_P$, and will not concern us here.
 
Supersymmetry would be a mathematical curiosity were it not for
the fact that it can be implemented consistently in local,
relativistic quantum field theory.  The supersymmetry charges,
$Q_\alpha$, can be obtained as Noether charges associated with
a conserved fermionic Noether current, $J^m_\alpha$,
\begin{eqnarray}
Q_\alpha & = & \int d^3x \; J^0_\alpha \nonumber \\
\partial_m J^m_\alpha & = & 0\;.
\end{eqnarray}

The simplest supersymmetric field theory is the Wess-Zumino model
\cite{wz model}, the
supersymmetric generalization of the toy model discussed above.  The
Wess-Zumino model involves one Weyl fermion, $\chi$, and two complex scalar
fields, $A$ and $F$.  The infinitesimal supersymmetry transformations
are as follows,
\begin{eqnarray}
\delta_\xi A &    = & ( \xi Q + \bxi \bQ) \times A \;=\; \sqrt{2} \,\xi\chi
\nonumber\\
\delta_\xi \chi & = & ( \xi Q + \bxi \bQ) \times \chi \;=\; i \sqrt{2}\, 
\sigma^m \bxi \partial_m A \;+\; \sqrt{2}\, \xi F  \nonumber\\
\delta_\xi F    & = & ( \xi Q + \bxi \bQ) \times F \;=\; i \sqrt{2}\, \bxi
\bar{\sigma}^m \partial_m \chi\;,
\label{WZ trans}
\end{eqnarray}
where $\xi$ is an anticommuting parameter.
It is a useful exercise to check that the supersymmetry transformations
close into the supersymmetry algebra,
\begin{equation}
[ \delta_\eta, \delta_\xi ]\, A \; = \; -\;2 i \,( \eta \sigma^m \bxi
- \xi \sigma^m \bbeta)\; \partial_m A\ ,
\end{equation}
and likewise for $\chi$ and $F$.

The Wess-Zumino model has the following Lagrangian \cite{wz model},
\begin{equation}
\L \; = \; \L_0 \; + \; \L_1\; ,
\end{equation}
where
\begin{equation}
\L_0 \; = \; -\;\partial_m A^* \partial^m A \;-\; i \bchi \bsigma^m
\partial_m \chi \;+\; F^*  F
\end{equation}
and
\begin{equation}
\L_1\;=\;M \,(A F \; - \; {1\over2}\; \chi\chi) \;+\;\lambda
\, (A^2 F - A \chi\chi) \; +\; {\rm h.c.}
\end{equation}
This Lagrangian is invariant
(up to a total derivative) under the supersymmetry transformations
(\ref{WZ trans}).

The equations of motion for $A,\; \chi$ and $F$ can be derived in
the usual way.  The fields $A$ and $\chi$ describe propagating,
physical particles.  The field $F$ does not propagate.  Its equation
of motion is algebraic,
\begin{equation}
{\partial \L \over \partial F} \; = \; F^* + M\, A + \lambda\, A^2\;=\;0
\;,
\label{F EOM}
\end{equation}
so $F$ can be eliminated using (\ref{F EOM}).  One finds
\begin{eqnarray}
\L & = & -\;\partial_m A^* \partial^m A\;- \;i \, 
\bchi\bsigma^m \partial_m \chi \;-\;\V(A^*,A) \nonumber\\
& & -\;{1\over2}\, M\, \chi\chi \;-\;{1\over2}\, M\, \bchi\bchi
\;-\;\lambda\, A\, \chi\chi \;-\;\lambda^* A^* \bchi\bchi \;,
\label{WZ Lag}
\end{eqnarray}
where the potential
\begin{equation}
\V(A^*,A) \; = \; |\, M\, A \; + \; \lambda\, A^2\, |^2
\end{equation}
is positive definite.

The Lagrangian (\ref{WZ Lag}) is the supersymmetric generalization
of the toy model discussed before.  It describes two physical
fields: one complex scalar and
one Weyl fermion, both of mass $M$.  The fields interact via Yukawa
and scalar couplings.  For the case at hand, $\lambda_F = \lambda$
and $\lambda_B = \lambda^* \lambda$.  These choices are fixed by
supersymmetry; they ensure that all quadratic divergences cancel
between bosonic and fermionic loops.

The equality of boson and fermion masses is a general feature of
supersymmetric field theories.  It follows from the fact that
$[P_m,\;Q_\alpha] = [P_m,\;\bQ_\dalpha] = 0$, which implies that
$P^2$ is a Casimir operator of the supersymmetry algebra.  The
absence of supersymmetric partners for the observed particles
means that supersymmetry must be broken in the everyday world.

\mysection{Coset Construction}

The Wess-Zumino model is instructive because it contains the
essential elements of supersymmetry.  However, it is just one
example of a supersymmetric field theory, and we would like
to be able to construct more at will.  In this section we will
develop a formalism which permits the construction of manifestly
supersymmetric quantum field theories.

In ordinary field theory, Poincar\'e symmetry is represented by
differential operators on scalar, spinor and vector fields.  Since
supersymmetry is a spacetime symmetry, it makes sense to represent
supersymmetry on superfields, supersymmetric generalizations of
ordinary fields.  The supersymmetry generators act as differential
operators on the superfields.

The systematic construction of superfields can be carried out
using a generalization of the coset construction of Callan,
Coleman, Wess and Zumino \cite{ccwz}, and Volkov \cite{volkov}.
The construction is rather involved, but it is so useful that we
will present it in complete generality \cite{VO}.  In the next
section we will specialize to the case of supersymmetry and superfields.

The coset construction proceeds as follows.  We start with a
group, $G$, of internal and spacetime symmetries, and partition
the (hermitian) generators of $G$ into the following three classes:
\begin{itemize}
\item
$\Gamma_A$, the generators of unbroken spacetime translations;
\item
$\Gamma_a$, the generators of spontaneously broken internal and
spacetime symmetries; and
\item
$\Gamma_i$, the generators of unbroken spacetime rotations
and unbroken internal symmetries.
\end{itemize}
The generators $\Gamma_i$ close into the stability group, $H$.

\begin{figure}[t]
\hspace*{1.1truein}
\psfig{figure=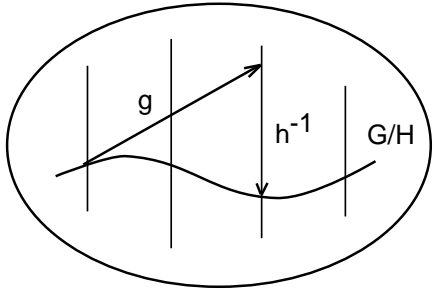,height=1.5in}
\caption{A schematic representation of the coset $G/H$.  The full space
represents the group $G$, while the vertical lines denote orbits under
$H$.  Note that a general $G$ transformation induces a compensating $H$
transformation to restore the section.}
\label{cosetpic}
\end{figure}

Given $G$ and $H$, we can construct the coset $G/H$.  We can
define the coset by an equivalence relation on the elements of
$G$,
\begin{equation}
\Omega \; \sim \; \Omega\, h\ ,
\label{coset}
\end{equation}
with $\Omega \in G$ and $h \in H$.  Therefore the coset can be pictured
as in Fig.~\ref{cosetpic}, as a section of a fiber bundle with total 
space, $G$, and fiber, $H$.

The definition (\ref{coset}) motivates us to parametrize the
coset as follows,
\begin{equation}
\Omega \; = \; e^{i X^A \Gamma_A} \; e^{i\pi^a(X) \Gamma_a}\;.
\end{equation}
Physically, the $X^A$ play the role of generalized spacetime
coordinates, while the $\pi^a(X)$ are generalized Goldstone
fields, defined on the generalized coordinates and valued in the
set of broken generators $\Gamma_a$.  There is one generalized
coordinate for every unbroken spacetime translation, and one
generalized Goldstone field for every spontaneously broken generator.

We define the action of the group $G$ on the coset $G/H$ by left
multiplication,
\begin{equation}
\Omega \; \rightarrow \;  g\, \Omega\; =\; \Omega^\prime\, h\; ,
\label{g trans}
\end{equation}
with $g \in G$.  In this expression
\begin{equation}
\Omega^\prime \; = \; e^{i X^{\prime A} \Gamma_A} \;
e^{i \pi^{\prime a}(X^\prime) \Gamma_a}
\end{equation}
and
\begin{equation}
h \; = \; e^{i \alpha^i (g, X, \pi) \Gamma_i}\ .
\label{define alpha}
\end{equation}

The group multiplication induces transformations
on the coordinates $X^A$ and the Goldstone fields $\pi^a$:
\begin{eqnarray}
X^A       & \rightarrow & X^{\prime A} \nonumber\\
\pi^a(X)  & \rightarrow & \pi^{\prime a}(X^\prime)\;.
\end{eqnarray}
These transformations realize the full symmetry group, $G$.  In the
general case, they are highly nonlinear functions of $g,\;\pi^a$
and $X^A$.  By construction, they linearize on the stability group,
$H$.  Furthermore, the field $\pi^a$ transforms by a shift under
the transformation generated by $\Gamma_a$.  This confirms that
$\pi^a$ is indeed the Goldstone field corresponding to the broken
generator $\Gamma_a$.

An arbitrary $G$ transformation induces a compensating $H$
transformation along the fiber, as shown in eq.~(\ref{g trans})
and Fig.~\ref{cosetpic}.  This transformation can be used
to lift any representation, $R$, of $H$, to a nonlinear
realization of the full group, $G$,
\begin{equation}
\psi(X) \; \rightarrow \psi^\prime(X^\prime) \; = \; D(h) \psi(X)\;.
\end{equation}
Here $D(h) = \exp(i \alpha^i T_i)$, where $\alpha^i$ was defined in
eq.~(\ref{define alpha}), and the $T_i$ are the hermitian generators
of $H$ in the representation $R$.

Having defined a nonlinear realization of $G$, we are now ready
to construct an invariant action.  The task is made easier by
identifying the vielbein, connection and covariant derivatives.
These are the covariant building blocks that we will use to
construct a $G$-invariant action.

The procedure is as follows.  We first construct the Maurer-Cartan
form, $\Omega^{-1} d\Omega$, where $d$ is the exterior derivative.  
The Maurer-Cartan form is valued in the Lie algebra of $G$, so it
has the expansion
\begin{equation}
\Omega^{-1} d \Omega \; = \; i (\omega^A \Gamma_A \; + \;
\omega^a \Gamma_a\; +\; \omega^i\Gamma_i)\;,
\label{omegas}
\end{equation}
where $\omega^A,\;\omega^a$ and $\omega^i$ are a set of
one-forms on the manifold parametrized by the coordinates $X^A$.

The Maurer-Cartan form transforms as follows under a rigid
$G$ transformation,
\begin{eqnarray}
\Omega & \rightarrow & g \; \Omega\; h^{-1} \nonumber \\
\Omega^{-1} d \Omega & \rightarrow & h (\Omega^{-1} d \Omega) h^{-1}
\;-\;dh\,h^{-1}\;.
\end{eqnarray}
Comparing with (\ref{omegas}), we see that $\omega^A$ and $\omega^a$
transform covariantly under $G$, while $\omega^i$ transforms by a
shift,
\begin{equation}
\omega\;\equiv\;\omega^i \Gamma_i 
\;\rightarrow\; h \omega h^{-1} + i\, dh\,h^{-1} \;.
\end{equation}
These transformations help us identify
\begin{equation}
\omega^A \; = \; dX^M \, E_M{}^A
\end{equation}
as the covariant vielbein,
\begin{equation}
\omega^a \; = \; dX^M \, E_M{}^A D_A \pi^a
\end{equation}
as the covariant derivative of the Goldstone field $\pi^a$, and
\begin{equation}
\omega^i \; = \; dX^M \, \omega_M^i
\label{h connection}
\end{equation}
as the connection associated with the stability group, $H$.

With these building blocks, it is easy to construct an
action invariant under the group $G$.  The first step is to
write all ordinary derivatives as covariant derivatives.
For the Goldstone fields, these are the $D_A \pi^a$ introduced
above.  For the others, they are
\begin{equation}
D_A \psi \;= \;E_A{}^M\,(\partial_M \, + \, \omega^i_M T_i )
\,\psi\ ,
\end{equation}
where $\omega^i_M$ is the $H$-connection (\ref{h connection}), and the
$T_i$ are the generators of $H$ in the representation $R$ of $\psi$.

Given the covariant derivatives, it is easy to write a $G$-invariant
action.  It is simply
\begin{equation}
S \; = \; \int d^D X\;\det E_M{}^A \; \L(\psi,\, D_A\psi,
\, D_A\pi^a)\;,
\label{general action}
\end{equation}
where $\L$ is a Lagrangian density, invariant under $H$.  The coset
construction ensures that the full action is automatically invariant
under $G$.

This construction is very general -- and very formal.  To see how
it works, let us consider the simplest possible case:  Poincar\'e
invariant field theory.  In this case, $G$ is the Poincar\'e group,
and $H$ its Lorentz subgroup.  There are no Goldstone fields, so we
identify
\begin{equation}
\Omega \; = \; e^{-iP_a x^a}\; ,
\end{equation}
where the $P_a$ are the usual momentum generators, and we replace $A$
by $a = 1,...,4.$

We will study the most general Poincar\'e transformation,
\begin{equation}
g \; = \; e^{i (P_a c^a + J^a{}_b \lambda^b{}_a)} \;,
\end{equation}
where the $J^a{}_b$ generate the Lorentz group, $H$.  The group
transformation
\begin{equation}
\Omega \; \rightarrow \; \Omega^{\prime} \; = \; g\, \Omega \, h^{-1}
\end{equation}
implies
\begin{equation}
x^{\prime a} \; = \; x^a \;- \; c^a\;+\;2\,\lambda^a{}_b x^b
\end{equation}
and
\begin{equation}
h \; = \; e^{i J^a{}_b \lambda^b{}_a}\;.
\end{equation}

By definition, a scalar field $\phi$ transforms as a singlet
under $H$,
\begin{equation}
\phi^\prime(x^\prime) \; = \; \phi(x)\ .
\label{def scalar}
\end{equation}
For an infinitesimal transformation, this reduces to
\begin{eqnarray}
\delta \phi(x) & = & \phi^\prime(x) \; - \; \phi(x) \nonumber \\[1mm]
& = & ( c^a \; - \; 2\, \lambda^a{}_b x^b)\, \partial_a \phi(x)\; .
\end{eqnarray}
A spinor field $\chi$ transforms as follows under $H$,
\begin{equation}
\chi^\prime(x^\prime) \; = \; D(h) \, \chi(x)
\end{equation}
where $D(h) = \exp (\lambda^{ab}\sigma^{ab})$.  
For infinitesimal transformations, this becomes
\begin{eqnarray}
\delta \chi(x) & = & ( c^a \; - \; 2\,\lambda^a{}_b x^b)\,
\partial_a \chi(x) \nonumber\\
& & +\; \lambda^{ab} \sigma^{ab}\, \chi(x)\;,
\end{eqnarray}
as expected for a spinor field.

To find the invariant Lagrangian, we construct the Maurer-Cartan form,
$\Omega^{-1} d\Omega = -i d x^a P_a$.  We extract the vielbein, $E_m{}^a
= \delta_m{}^a$, and the connection, $\omega^i = 0$.  We see that the
covariant derivative $D_a$ is just $\partial_a$.  With these results, we
are able to construct a Poincar\'e invariant action.  We find
\begin{equation}
S \; = \; \int d^4x\; \L (\phi,\,\partial_a \phi,\,\psi,
\,\partial_a\psi)\;,
\label{poincare action}
\end{equation}
where the Lagrangian density, $\L$, is invariant under the Lorentz
group, $H$.  Equation (\ref{poincare action}) is nothing
but the usual Poincar\'e invariant action for quantum field theory
-- derived in the most sophisticated possible way!

\mysection{General Superfields}

The coset construction is much too technical for the case of ordinary
Poincar\'e-invariant field theory.  It just reproduces what we
already know.  However, for the case of supersymmetry,
the coset construction leads to something new:  a manifestly
supersymmetric technique for constructing supersymmetric quantum
field theories \cite{mybook,superf}.

In this section we shall see how this works.  We will take $G$ to
be the supergroup generated by the supersymmetry algebra (\ref{susy
alg}).  We take the group $H$ to be the Lorentz group, and we choose
to keep all of $G$ unbroken.  Therefore we have
\begin{equation}
\Omega \; = \; e^{i (-x^a P_a + \theta^\alpha Q_\alpha +
\btheta_\dalpha \bQ^\dalpha)}\;,
\end{equation}
where the generalized spacetime coordinates are $z = (x,\,
\theta,\,\btheta)$.  The coordinates $\theta$ and $\btheta$
are Lorentz spinors, so we take them to anticommute
\begin{equation}
\{\theta^\alpha,\;\theta^\beta\}\;=
\;\{\btheta_\dalpha,\;\btheta_\dbeta\}\;=
\;\{\theta^\alpha,\;\btheta_\dbeta\}\;=\;0\;.
\end{equation}
We call the coordinates $(x,\,\theta,\,\btheta)$ superspace.

A supersymmetry transformation is specified by the group element
\begin{equation}
g \; = \; e^{i ( \xi^\alpha Q_\alpha + \bxi_\dalpha \bQ^\dalpha)}\;,
\end{equation}
with anticommuting parameters $(\xi,\;\bxi)$.  The transformation
\begin{equation}
\Omega \; \rightarrow \; \Omega^\prime \; = \; g \Omega h^{-1}
\end{equation}
induces the motion
\begin{eqnarray}
x^a & \rightarrow & x^a \; + \; i \theta \sigma^a \bxi\; -\; i 
\xi \sigma^a \btheta \nonumber \\
\theta^\alpha & \rightarrow & \theta^\alpha \; + \; \xi^\alpha
\nonumber \\
\btheta_\dalpha & \rightarrow & \btheta_\dalpha\; +\;
\bxi_\dalpha
\end{eqnarray}
and
\begin{equation}
h \; = \; 1\;.
\end{equation}

Given these transformations, we define a scalar superfield 
$F(z)$ in analogy to (\ref{def scalar}),
\begin{equation}
F^\prime(z^\prime) \; = \; F(z)\;.
\end{equation}
Under an infinitesimal supersymmetry transformation, this
reduces to
\begin{eqnarray}
\delta_\xi F(z) & = & F^\prime(z)\;-\;F(z) \nonumber \\[1mm]
& = & ( \xi^\alpha Q_\alpha \; + \; \bxi_\dalpha\bQ^\dalpha)
\,F(z)\; ,
\end{eqnarray}
where the differential operators $Q$ and $\bQ$ are
\begin{eqnarray}
Q_\alpha & = & {\partial \over \partial \theta^\alpha}\;-
\;i \sigma^m_{\alpha \dalpha} \btheta^\dalpha\, \partial_m
\nonumber \\
\bQ^\dalpha & = & {\partial \over \partial \btheta_\dalpha}\;-
\;i \theta^\alpha \sigma^m_{\alpha \dbeta} \epsilon^{\dbeta \dalpha}
\,\partial_m\;.
\label{susy Q}
\end{eqnarray}
The anticommuting derivatives obey the relations
\begin{eqnarray}
{\partial \over \partial \theta^\alpha}\, \theta^\beta & = &
\delta_\alpha^\beta \nonumber \\
{\partial \over \partial \theta^\alpha}\, \theta^\beta \theta^\gamma 
& = & \delta_\alpha^\beta \theta^\gamma\;-\;\theta^\beta 
\delta_\alpha^\gamma \;,
\end{eqnarray}
and similarly for $\btheta$.  It is a useful exercise to check
the differential operators $Q$ and $\bQ$ close into the
supersymmetry algebra:
\begin{eqnarray}
\{ Q_\alpha, \;\bQ_\dalpha\} & = & 2i
\sigma_{\alpha \dalpha}^m\, \partial_m \nonumber\\
\{ Q_\alpha, \; Q_\beta \} & = & 
\{ \bQ_\dalpha, \;\bQ_\dbeta\} \; = \; 0 \;.
\label{Q alg}
\end{eqnarray}
This ensures that superfields do indeed represent the supersymmetry
algebra.

To find an invariant action, we compute the Maurer-Cartan form,
$\Omega^{-1} d\Omega$.  It is a useful exercise to extract the
vielbein,
\begin{equation}
E_M{}^A \; = \; \pmatrix{
                     \delta_m{}^a & 0 & 0 \crr
                     - i \sigma^a_{\mu\dmu} \btheta^\dmu &
                                \delta_\mu{}^\alpha & 0\crr
                     - i \theta^\mu \sigma^a_{\mu\dnu} 
                                \epsilon^{\dnu\dmu} & 0 &
                                \delta^\dmu{}_\dalpha \crr }\ ,
\end{equation}
and the $H$-connection,
\begin{equation}
\omega^i \; = \; 0\;.
\end{equation}
Then the covariant derivative of a scalar superfield is just
\begin{equation}
D_A F(z) \; = \; E_A{}^M\, \partial_M F(z)\;,
\end{equation}
where the supersymmetric covariant derivatives are
\begin{eqnarray}
D_a & = & \partial_a \nonumber \\[1mm]
D_\alpha & = & {\partial \over \partial \theta^\alpha} \; +
\; i \sigma^m_{\alpha \dalpha}\btheta^\dalpha \, \partial_m \nonumber \\
\bD^\dalpha & = & {\partial \over \partial \btheta_\dalpha} \; + 
\; i\theta^\alpha \sigma^m_{\alpha \dbeta} \epsilon^{\dbeta \dalpha} \,
\partial_m\;.
\label{susy D}
\end{eqnarray}

By construction, the supersymmetric covariant derivatives
(anti)commute with the supersymmetry generators,
\begin{equation}
\{ Q_\alpha, \; D_\beta \} \; =
\; \{ \bQ_\dalpha, \; \bD_\dbeta \} \; =
\; \{ Q_\alpha, \; \bD_\dbeta \} \; =
\; \{ \bQ_\dalpha, \; D_\beta \} \; =\; 0\;.
\end{equation}
They also obey the following structure relations
\begin{eqnarray}
\{ D_\alpha, \; \bD_\dalpha \} & = & - 2 i \,
\sigma^m_{\alpha \dalpha} \, \partial_m \nonumber \\
\{ D_\alpha, \; D_\beta \} & = & \{ \bD_\dalpha, \;
\bD_\dbeta \} \; = \;  0\;.
\end{eqnarray}

To make contact with physics, we must extract $x$-dependent
component fields from the superfields.  This can be done by
expanding the superfields in terms of $\theta$ and $\btheta$:
\begin{eqnarray}
F(x,\theta,\btheta) & = & f(x) \; + \; \theta\varphi(x) \; 
+ \; \btheta\bphi(x)  \nonumber \\[1mm]
&&  + \; \theta\theta\, m(x) \; +
\; \btheta\btheta\, n(x)
\ + \; \theta\sigma^m\btheta v_m(x)  \nonumber \\[1mm]
&& +\ \theta\theta\btheta\blambda(x)
\ +\ \btheta\btheta\theta\psi(x)
\ + \; \theta\theta\btheta\btheta \, d(x)\;.
\label{general scalar}
\end{eqnarray}
The expansion terminates because $\theta$ and $\btheta$
anticommute.  This implies that a given superfield contains a
finite number of component fields.

The supersymmetry transformations of the component fields can be
found from the supersymmetry transformations of the superfields,
\begin{eqnarray}
\delta_\xi F(x,\theta,\btheta) & = &
(\xi Q\ +\ \bxi \bQ)\,F(x,\theta,\btheta) \nonumber \\[1mm]
& = & \delta_\xi f(x) \; + \; 
\theta\,\delta_\xi\varphi(x) \; + \; \btheta\,
\delta_\xi\bphi(x) \nonumber \\[1mm]
&&  + \; \theta\theta\, \delta_\xi m(x) \; + \; 
\btheta\btheta\, \delta_\xi n(x)
\ + \; \theta\sigma^m\btheta\, \delta_\xi v_m(x)  \nonumber \\[1mm]
&& +\ \theta\theta\btheta\,\delta_\xi \blambda(x)
\ +\ \btheta\btheta\theta\,\delta_\xi \psi(x)
+ \; \theta\theta\btheta\btheta \, \delta_\xi d(x)\;.
\end{eqnarray}
By construction, the component transformations close into
supersymmetry algebra.

\mysection{Chiral Superfields}

In this section we will write the Wess-Zumino model \cite{wz model}
in manifestly covariant form.  Our results will serve as the first
step towards constructing more general supersymmetric theories with
spin-zero and spin-$\frac{1}{2}$ fields.

At first glance, it might seem simple to write down the Wess-Zumino
model in terms of the superfields discussed in the previous section.
However, the problem is harder than it first appears because
a general scalar superfield contains far too many component fields.
We must first reduce the number of component fields by imposing a
covariant constraint.

It turns out that the right constraint is just \cite{mybook}
\begin{equation}
\bD_\dalpha \Phi \; = \; 0\;.
\label{dphi}
\end{equation}
This defines the chiral superfield, $\Phi$.  The constraint is
consistent in the sense that it is covariant, and does not impose
equations of motion on the component fields.

We can solve the constraint (\ref{dphi}) by writing $\Phi$ as
a function of $y$ and $\theta$, where
\begin{equation}
y^m \; = \; x^m \; + \; i \theta \sigma^m \btheta\;.
\end{equation}
Since $\bD \theta = \bD y= 0$, the field $\Phi(y,\theta)$
automatically satisfies the constraint (\ref{dphi}).

To find the component fields, we expand $\Phi(y,\theta)$ in
terms of $\theta$,
\begin{eqnarray}
\Phi(y, \theta) & = & A(y) \; + \; \sqrt{2} \, \theta \chi(y) \;
+ \; \theta\theta \, F(y) \nonumber \\[2mm]
& = & A(x) \; + \; i \theta \sigma^m \btheta \, \partial_m A(x) \; 
+ \;{1\over4}\, \theta \theta \btheta \btheta \Box A(x) \nonumber \\
&& + \; \sqrt{2} \, \theta \chi(x) \; - \; {i\over\sqrt 2}\, \theta\theta \,
\partial_m \chi(x) \sigma^m \btheta \; + \; \theta \theta \, F(x)\;.
\label{phi expansion}
\end{eqnarray}
Equation (\ref{phi expansion}) shows that the chiral superfield $\Phi$
contains the same component fields as the Wess-Zumino model.

The supersymmetry transformations of the component fields can
be found using the differential operators (\ref{susy Q}),
\begin{eqnarray}
\delta_\xi \Phi & = & (\xi Q \; + \; \bxi \bQ ) \, \Phi \nonumber \\[1mm]
& = & \delta_\xi A(x) \; + \; \sqrt{2} \, \theta\, \delta_\xi\chi(x) \; + \; 
\theta \theta \, \delta_\xi F(x)\;+\;....
\end{eqnarray}
This gives
\begin{eqnarray}
\delta_\xi A & = & \sqrt{2} \,\xi\chi \nonumber\\
\delta_\xi \chi & = & i \sqrt{2}\, \sigma^m\bxi\, \partial_m A 
\;+\;\sqrt{2}\, \xi F  \nonumber\\
\delta_\xi F & = & i \sqrt{2}\, \bxi
\bsigma^m \partial_m \chi\;.
\end{eqnarray}
in accord with (\ref{WZ trans}).

Now that we have the chiral superfield $\Phi$, we can construct
a supersymmetric action.  With superfields, the task is trivial.  
According to (\ref{general action}), an invariant action is just
\begin{equation}
S\;=\;\int d^4x d^4\theta \; \det\,E_M{}^A \; \L(\Phi,\,
D_A\Phi)\;.
\label{susy action}
\end{equation}
For the case at hand, $\det\,E_M{}^A = 1$, so (\ref{susy action})
reduces to
\begin{equation}
S\;=\;\int \; d^4x d^4\theta \; \L(\Phi,\, D_A\Phi)\;.
\label{susyact}
\end{equation}
This can be expressed in terms of ordinary fields using the fact
that 
\begin{equation}
\int d^4 \theta \; \equiv \; {\partial^2 \over \partial
\theta^2} \; {\partial^2 \over \partial \btheta^2}\;.
\end{equation}
By construction, the action (\ref{susyact}) is manifestly
supersymmetric.

To check that (\ref{susyact}) is indeed invariant, note that
$\L$ is itself a superfield, so $\delta \L = (\xi Q  + \bxi \bQ) \L$.
From the form of the differential operators $Q$ and $\bQ$, it is not
hard to see that the $\theta \theta \btheta \btheta$ component of any
superfield transforms into a total derivative.  Since the action
(\ref{susyact}) is a spatial integral, it is automatically invariant
under supersymmetry.

The form of the Lagrangian can be found by dimensional analysis.
For the action to be dimensionless, the Lagrangian must have
dimension two.  There are just two possible choices:  $\Phi^+ \Phi$
and $\Phi^2$.  The integral of $\Phi^2$ is zero, so $\Phi^+ \Phi$ is
the only possible term.

To confirm that $\Phi^+ \Phi$ is the superspace Lagrangian, we can use
the expansion (\ref{phi expansion}) to write $\Phi^+ \Phi$
in terms of component fields.  We find
\begin{eqnarray}
\Phi^+\Phi & = & A^*A \;+\;...\;+\;\theta\theta \btheta \btheta\;
\bigg[\, {1\over4}\,  A^* \Box A \; + \; {1\over4}\, \Box A^* A \; 
- \;{1\over2}\, \partial_m A^* \partial^m A \nonumber \\
& &  \;+\;F^* F \; +\; {i\over2} \, \partial_m \bchi \bsigma^m
\chi \; - \;{i\over2} \, \bchi \bsigma^m \partial_m \chi \, \bigg] .
\end{eqnarray}
This shows that
\begin{equation}
\int d^4x d^4\theta \; \Phi^+\Phi
\end{equation}
is indeed the supersymmetric kinetic energy for the Wess-Zumino
model.

To recover the full Wess-Zumino model, we also need superspace
expressions for the masses and couplings.  We will take advantage
of the fact that for chiral superfields,
\begin{equation}
\int d^4x d^2\theta \; \Phi(x,\theta,\btheta)\;=
\;\int d^4x d^2\theta \; \Phi(x,\theta)
\label{chiral invt}
\end{equation}
is also a supersymmetry invariant.  It is not hard to check that
(\ref{chiral invt}) is actually supersymmetric.  This can
be seen from first principles, using $\delta\Phi = (\xi Q + \bxi\bQ)
\Phi$, or from the component transformation law for $F$, the
$\theta\theta$ component of the chiral superfield, $\Phi$.

Since the product of any two chiral superfields is also a chiral
superfield, eq.~(\ref{chiral invt}) can be used to construct
renormalizable supersymmetric interactions for chiral superfields.  
The invariant action is just
\begin{equation}
S \;=\;\int \; d^4x d^2\theta \; P(\Phi)\ ,
\end{equation}
where the superpotential, $P(\Phi)$, is analytic in $\Phi$.  By power
counting, we see that renormalizability requires the superpotential
to have degree at most three.  Therefore
\begin{equation}
P(\Phi) \; = \;  {1\over2}\,m\,\Phi^2 \; + \; {1\over3}\, 
\lambda\, \Phi^3
\end{equation}
is the most general renormalizable interaction for a single chiral
superfield.  (A linear term can be eliminated by a shift.)

The superpotential characterizes the interactions of chiral
superfields.  Indeed, it gives rise to
\begin{itemize}
\item
Fermion masses and Yukawa couplings,
\begin{equation}
{\partial^2 P \over \partial A^2} \; \chi\chi\ ,
\end{equation}
\item
The scalar potential,
\begin{equation}
\V (A,\;A^*) \; = \; \Big| \, {\partial P \over \partial A}\,
\Big|^2\;.
\end{equation}
\end{itemize}
These expressions follow from the auxiliary field equation of
motion,
\begin{equation}
F^*\ +\ {\partial P\over\partial A}\ =\ 0\;.
\end{equation}

\mysection{Vector Superfields}

In the previous section we found that chiral superfields
describe supersymmetric matter fields with spins zero and
$\frac{1}{2}$.  In this section we will construct the
supersymmetric extensions of ordinary spin-one gauge fields.

We will start by studying the gauge transformations of chiral
superfields.  We assume that under a rigid symmetry transformation, 
$\Phi$ transforms in a representation, $R$, of an (unbroken)
internal symmetry group,
\begin{equation}
\Phi \; \rightarrow \; e^{i \alpha^{(a)} T^{(a)}} \; \Phi\ ,
\label{int trans}
\end{equation}
where the $T^{(a)}$ are the hermitian generators of the group in
the representation $R$.  Our goal is to gauge this symmetry by
making $\alpha$ local while preserving the constraint $\bD
\Phi = 0$. This requires that we promote $\alpha$ to a chiral
superfield, $\Lambda$, with $\bD \Lambda = 0$.  Then
\begin{equation}
\Phi \; \rightarrow \; e^{i \Lambda^{(a)} T^{(a)}}\; \Phi
\end{equation}
is a fully supersymmetric local symmetry transformation.

Let us assume that the supersymmetric action
\begin{equation}
S\ =\ \int d^4x d^4\theta \; \Phi^+ \Phi \; + \; \Bigg[ \int d^4x d^2 
\theta \; P(\Phi) \; + \; {\rm h.c.} \Bigg]
\end{equation}
is invariant under the rigid transformation (\ref{int trans}).  This
requires that the superpotential $P(\Phi)$ be invariant under the
internal symmetry group.  Now let $\alpha$ be lifted to
$\Lambda$.  The superpotential is still invariant.  The
kinetic term, however, is not,
\begin{equation}
\Phi^+ \Phi \; \rightarrow \; \Phi^+ \; e^{-i \Lambda^+} \;
e^{i \Lambda}\; \Phi\ ,
\end{equation}
where $\Lambda = \Lambda^{(a)} T^{(a)} $.  

The kinetic term can be made invariant by introducing a vector
superfield, $V = V^{(a)} T^{(a)}$, with
\begin{equation}
V^+ \; = \; V\;,
\end{equation}
such that
\begin{equation}
e^{gV} \; \rightarrow \; e^{i\Lambda^+} \, e^{gV} \, 
e^{-i\Lambda}
\label{gauge trans}
\end{equation}
under a gauge transformation.  In this way
\begin{equation}
S\ =\ \int d^4x d^4\theta \; \Phi^+ e^{gV} \Phi \; + \; \Bigg[
\int d^4x d^2  \theta \; P(\Phi) \; + \; {\rm h.c.} \Bigg]
\label{gauge Lagrangian}
\end{equation}
is a supersymmetric and gauge invariant action.

The vector field $V$ contains many component fields, which we write
in the following form \cite{mybook},
\begin{eqnarray}
V & = & C(x) \; + \; i \theta\eta(x) \; - i \btheta \bbeta(x) 
\; -\;\theta \sigma^m \btheta v_m(x) \nonumber\\
&  &+ \;{i\over2} \, \theta\theta \bigg(M(x) + i N(x)\bigg) \; - \; {i\over2}\,
\btheta\btheta \bigg(M(x) - i N(x)\bigg) \nonumber \\
& & + \; i \theta\theta \btheta
\bigg(\blambda(x) + {i\over2} \bsigma^m \partial_m \eta(x)\bigg) \;-\; i
\btheta\btheta\theta \bigg(\lambda(x)  + {i\over2} \sigma_m \partial_m
\bbeta(x)\bigg) \nonumber \\
& & + \;{1\over2} \, \theta \theta \btheta \btheta  \bigg(D(x)  + \frac{1}{2}
\Box C(x)\bigg)\ .
\label{v expansion}
\end{eqnarray}
However, half are gauge degrees of freedom.  To see this,
note that under a gauge transformation,
\begin{equation}
gV \; \rightarrow \; gV \; -\; i \,(\Lambda - \Lambda^+) \; + \; ...
\end{equation}
where
\begin{eqnarray}
i \,(\Lambda - \Lambda^+) & = & i\, (A  -  A^*) \; + \; i \sqrt{2} 
(\theta \chi - \btheta \bchi) \;+\; i \theta \theta F \;
- \; i \btheta \btheta F^* \nonumber \\
&&\;-\; \frac{1}{\sqrt{2}}\, \theta \theta \btheta \bsigma^m 
\partial_m \chi 
\;+\;\frac{1}{\sqrt{2}} \, \btheta \btheta \theta \sigma^m
\partial_m \bchi \nonumber \\
&&\;-\; \theta \sigma^m \btheta\, \partial_m (A + A^*)\;+
\; \frac{i}{4} \theta \theta \btheta \btheta \, \Box (A - A^*)\ .
\label{lambda expansion}
\end{eqnarray}
Comparing (\ref{v expansion}) with (\ref{lambda expansion}), we see that
$C,\;\eta,\; M$ and $N$ can all be gauged away,
\begin{equation}
C\ =\ \eta\ =\ M\ =\ N\ =\ 0\ .
\label{wz gauge}
\end{equation}
The component field $v_m$ still transforms  as
\begin{equation}
v_m \; \rightarrow \; v_m \; - \; \partial_m \alpha\ ,
\end{equation}
where $\alpha \equiv 2\,{\rm Re}\, A$.

Equation (\ref{wz gauge}) defines the Wess-Zumino gauge.  In this
gauge the vector superfield $V$ takes a simple form,
\begin{equation}
V \;= \; - \; \theta \sigma^m \btheta v_m
\;-\;i \btheta \btheta \theta \lambda \; +\; i \theta \theta
\btheta \blambda
\;+\; \frac{1}{2}  \, \theta \theta \btheta \btheta \; D\;.
\end{equation}
A vector superfield contains just the
right components to be the supersymmetric generalization of a vector
field.  It has a spin-one vector boson and its spin-$\frac{1}{2}$
fermionic partner.  The real scalar $D$ is an auxiliary field.

Equation (\ref{gauge Lagrangian}) gives rise to gauge-invariant
kinetic terms for all the matter fields.  We also need kinetic
terms for the gauge fields themselves.  In particular, we need to
find a superfield generalization of the covariant field strength,
$F_{mn}$.  It is
\begin{eqnarray}
W_\alpha & \equiv &  - \; {1\over4 g} \, \bD\bD \, e^{-gV} D_\alpha
\, e^{gV} \nonumber \\
& & = \; - \; {1\over 4} \, \bD\bD \, D_\alpha V \; + \; ....
\end{eqnarray}
By construction, $W_\alpha$ is a chiral superfield.  It is also
gauge-covariant,
\begin{equation}
W_\alpha \; \rightarrow \; e^{i\Lambda} \; W_{\alpha} \;
e^{-i\Lambda}\ ,
\end{equation}
under a gauge transformation (\ref{gauge trans}).  In abelian case,
it is easy to check that
\begin{eqnarray}
W_\alpha & = & - \;  {1\over4} \, \bD \bD D_\alpha V \nonumber \\
& \rightarrow & - \; {1\over4} \, \bD \bD D_\alpha \, (V - i\Lambda
+ i\Lambda^+) \nonumber \\
& &=\;W_\alpha \; + \; {i \over4} \, \bD \bD D_\alpha \Lambda
\nonumber \\
& &=\;W_\alpha \; + \; {i \over4} \, \bD \{ \bD, D_\alpha\} \Lambda
\nonumber \\
& &=\;W_\alpha\;,
\end{eqnarray}
where we have set $g=1$ for simplicity.

In terms of component fields, we see that $W_\alpha$ is indeed the
supersymmetric generalization of the field strength $F_{mn}$.  It has
the following $\theta$-expansion,
\begin{equation}
W_\alpha \;=\; - \;i \, \lambda_\alpha \; + \; [ \delta_\alpha{}^\beta \, 
D \; -\;i \, \sigma_{mn\alpha}{}^\beta \, F_{mn} ] \; \theta_\beta
\;+\; \theta \theta \, \sigma^m_{\alpha \dalpha} \, \D_m
\blambda^\dalpha\;+\ ...\ ,
\end{equation}
where $\D_m\blambda$ is the gauge-covariant derivative of $\blambda$.  

We now have what we need to construct the most general renormalizable
action involving gauge and matter fields.  It is
\begin{eqnarray}
S & = & \int d^4x d^4\theta \; \Phi^+ e^{gV} \Phi \nonumber \\
& + &  \bigg[\,\int  d^4x d^2\theta \; \bigg(\,{1\over4}\,W^{(a)}
W^{(a)} \;+\;P(\Phi)\,\bigg) \; + \; {\rm h.c.}\,\bigg]\ ,
\end{eqnarray}
where the superpotential is gauge-invariant and analytic of degree
at most three.  All the terms are fixed by symmetry -- except for those
in the superpotential.

\renewcommand{\arraystretch}{1.6}
\begin{table}[t]
\caption{The Vector Superfields of the MSSM.}
\begin{center}
\vspace{0.4cm}
\begin{tabular} {|c|c|c|c|} \hline
Superfield &\ \ SU(3) $\times$ SU(2) $\times$ U(1)\ 
\ &\ \ Particles\ \ \\\hline
$V^a$ & (8, 1, 0) & gluons and gluinos ($\tilde g$) \\
$V^i$ & (1, 3, 0) & $W$'s and winos ($\tilde W$) \\
$V$ & (1, 1, 0) & $B$ and bino ($\tilde B$) \\\hline
\end{tabular}
\end{center}
\end{table}
\renewcommand{\arraystretch}{1.0}

The component Lagrangian can be found by eliminating the auxiliary
fields, $F$ and $D$.  It is simply
\begin{eqnarray}
\L & = & -\ \D_m A^*_i \D^m A^i \; - \; i \bchi_i \bsigma^m  \D_m \chi^i
\;-\;{1\over4} \, F_{mn}^{(a)} F^{mn(a)} \;-\;i \blambda^{(a)} 
\bsigma^m \D_m \lambda^{(a)} \nonumber \\
&& -\;i  \sqrt{2}\, g \, \blambda^{(a)} \bchi_i \, T^{(a)i}{}_j \, A^j
\;+\;i  \sqrt{2}\, g \,  A^*_j\, T^{(a)j}{}_i \, \chi^i \lambda^{(a)}
\nonumber \\
&& -\;{1\over2} \, P_{ij}\, \chi^i \chi^j \; - \; {1\over2} \, (P_{ij})^* \,
\bchi_i \bchi_j
\;-\;| \, P_i \, |^2 \; - \; {1\over2} \, g^2 \, D^{(a)2}\ ,
\end{eqnarray}
where all derivatives are gauge covariant, and we have explicitly
labeled the matter fields by an index $i,j,\;....$  In this
expression,
\begin{eqnarray}
P_{ij} & = & {\partial^2 \over \partial A^i \partial A^j}\,P(A) \nonumber \\[2mm]
P_{i} & = & {\partial \over \partial A^i }\,P(A)
\end{eqnarray}
and
\begin{equation}
D^{(a)} \;= \;A^*_j \, T^{(a)j}{}_i \, A^i
\end{equation}

\mysection{The Supersymmetric Standard Model}

We now have the tools we need to construct the MSSM -- the minimal
supersymmetric version of the standard model \cite{earlyfayet} --
\cite{susysu5b}.  (See also \cite{ferrarabook} --
\cite{ssm}.)  We will start by
defining the superfield content of the model.  We will then write
down the supersymmetric Lagrangian and study its implications.

The MSSM is based on the same SU(3) $\times$ SU(2) $\times$ U(1) gauge
group as the ordinary standard model.  Therefore it requires a color
octet of vector superfields $V^{(a)}$, as well as a weak triplet $V^{(i)}$ 
and a hypercharge singlet $V$.  These superfields contain the appropriate
spin-one gauge bosons, as well as their spin-$\frac{1}{2}$ partners, as
shown in Table 1.

The vector superfields interact with the superfield versions of the
quarks and the leptons.  These superfields are shown in Table 2.  They
are chiral superfields; they contain the spin-$\frac{1}{2}$ quarks and
leptons, as well as their spin-zero partners, the squarks and sleptons.

\renewcommand{\arraystretch}{1.6}
\begin{table}[t]
\caption{The Chiral Superfields of the MSSM.}
\begin{center}
\vspace{0.4cm}
\begin{tabular} {|c|c|c|c|} \hline
Superfield &\ \ SU(3) $\times$ SU(2) $\times$ U(1)\ 
\ &\ \ Particles\ \ \\\hline
$Q$ & (3, 2, 1/6) & quarks $(u,d)$ and squarks $(\tilde u,\tilde d)$\\
$\bU$ & ($\bar{3}$, 1, $-2/3$) & quarks $(\bar{u})$ and squarks 
($\tilde{\bar{u}}$)\\
$\bD$ & ($\bar{3}$, 1, 1/3) & quarks $(\bar{d})$ and squarks
($\tilde{\bar{d}}$)\\
$L$ &   (1, 2, $-1/2$) & leptons $(\nu,e)$ and sleptons $(\tilde \nu, \tilde e)$\\
$\bE$ & (1, 1, 1) & electron ($\bar e$) and selectron $(\tilde{\bar{e}}$) 
\\\hline
$H_1$ & (1, 2, $-1/2$) & Higgs ($h_1$) and Higgsinos ($\tilde H_1$) \\
$H_2$ & (1, 2, 1/2) & Higgs ($h_2$) and Higgsinos ($\tilde{H_2}$)
\\\hline
\end{tabular}
\end{center}
\end{table}
\renewcommand{\arraystretch}{1.0}

The supersymmetric extensions of Higgs bosons are also shown
in Table 2.  They include two complex Higgs doublets, 
$(h_1,h_2)$, as well as their spin-$\frac{1}{2}$ partners, the two
Higgsinos.   In supersymmetric theories, two (or more) Higgs doublets
are required for the Higgsino anomalies to cancel among themselves.

When the gauge symmetry is broken, three of the scalar Higgs particles
that are eaten by the $W$ and $Z$.  The remaining
five scalars include two neutral CP-even bosons, $h$ and $H^0$, one
charged boson $H^\pm$, and one neutral CP-odd boson $A$.

The spin-$\frac{1}{2}$ Higginos mix with the winos and binos.  The mass
eigenstates include four neutral two-component spinors, $\chi^0_i,$ with
$i = 1,...,4$, and two charged spinors, $\chi^\pm_i$, $i = 1, 2$.
These particles are called neutralinos and charginos, respectively.

The kinetic terms of all the fields are fixed by gauge invariance.
They are simply
\begin{eqnarray}
\L & = & \int d^4\theta\; \Phi^+ \exp\Big(\,g_1 V T
+ g_2 V^{(i)} T^{(i)} + g_3 V^{(a)} T^{(a)} \,\Big) \,\Phi\nonumber \\[1mm]
& & + \; \bigg[\,\int d^2\theta \; {1\over4}\,\bigg(\, W^{(a)} W^{(a)}
\;+\ W^{(i)} W^{(i)} \;+\; WW\, \bigg) \; + \; 
{\rm h.c.} \, \bigg] \ ,
\label{MSSM KE}
\end{eqnarray}
where $\Phi$ is a vector of the matter superfields, $\Phi = (Q,\,
\bU,\,\bD,\,L,\,\bE,\,H_1,\,H_2)^T$, and the generators $(T^{(a)},\,
T^{(i)},\,T)$ are chosen to be in the appropriate representations of the
SU(3) $\times$ SU(2) $\times$ U(1) gauge group.  The Lagrangian
(\ref{MSSM KE}) contains the gauge couplings $g_3,\, g_2$ and $g_1$.  
They obey the standard-model relation, $e = g_1 \cos\theta$, where
$\cos^2 \theta = g^2_2/(g_1^2 + g_2^2)$ is the usual weak mixing
angle.

\begin{figure}[t]
\hspace*{0.5truein}
\psfig{figure=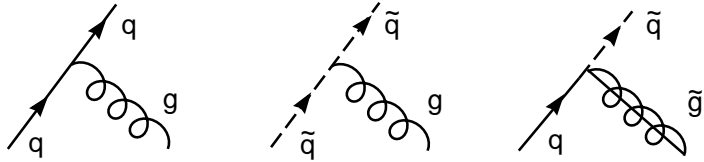,height=1.0in}
\caption{Some of the vertices that arise from the supersymmetric
kinetic terms.  All these vertices are proportional to the strong
coupling $g_3$.  The first two are ordinary gauge couplings, but
the third is a Yukawa coupling.  The Yukawa is necessary
to cancel quadratic divergences induced by gauge boson loops.}
\label{ke diags}
\end{figure}

The matter fields interact with the vector fields by supersymmetric
generalizations of the ordinary gauge interactions.  Some sample 
vertices are shown in Fig.~\ref{ke diags}.  For each such vertex,
the strength of the interaction is fixed by the appropriate gauge
coupling.  Note that in each vertex, superparticle number is
conserved, modulo two.

The Yukawa couplings and scalar potential are defined by the
superpotential, $P$.  For the case at hand, the most general
renormalizable gauge-invariant superpotential is just
\begin{eqnarray}
P & = & \mu\, H_1 H_2 \; + \; \lambda_U \, Q \bU H_2
\;+\; \lambda_D \, Q \bD H_1 \; + \; \lambda_E \, L \bE H_1
\nonumber \\[2mm]
& &+\;\left\{ \, L H_2 \; + \; Q L \bD \;+
\;\bU \bD \bD \; + \; L L \bE \right\}\ .
\label{big super}
\end{eqnarray}
Here $\lambda_U, \; \lambda_D$ and $\lambda_E$ are the usual quark
and lepton Yukawa matrices, and $\mu$ is the supersymmetric Higgs
mass parameter.  (We have suppressed a sum over generations.)

\begin{figure}[t]
\hspace*{1.0truein}
\psfig{figure=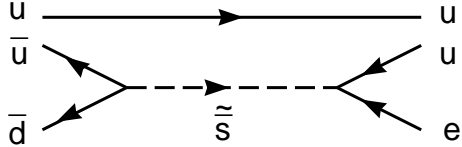,height=1.0in}
\caption{A diagram that contributes to squark-mediated proton decay.}
\label{pdecay}
\end{figure}

The terms in brackets are a striking feature of the MSSM.  They
give rise to dimension-four operators which violate baryon
and lepton number and lead to instantaneous proton decay
\cite{pdecaypapers}, as
shown in Fig.~\ref{pdecay}.  The fact that dimension-four
operators violate $B$ and $L$ contrasts sharply with the
ordinary standard model, where $B$ and $L$ violation first
appears at dimension six.

Clearly, for the MSSM to be phenomenologically viable, these
operators must be suppressed.  One way to accomplish this is to tune
their coefficients to be acceptably small.  Another is to eliminate
them entirely.  We shall take the second approach, and set all
these terms to zero.

Of course, the only natural way to eliminate these operators is to
impose a symmetry to forbid them.  We could always {\it impose}
baryon and lepton number conservation, but that would completely
forbid proton decay.  It turns out that there is another
symmetry, known as $R$-parity \cite{r parity}, which
eliminates the renormalizable
terms, but still allows proton decay via higher-dimensional
operators 

$R$-parity is a $Z_2$ symmetry under which the
vector superfields remain invariant,
\begin{equation}
\left\{\begin{array}{c}
V^{(a)} \\ V^{(i)} \\ V \end{array}\right\}
\; \rightarrow \;
\left\{ \begin{array}{c}
V^{(a)} \\ V^{(i)} \\ V \end{array}\right\}
\end{equation}
while the coordinates $\theta \rightarrow  - \theta$.  The chiral
superfields transform as follows:
\begin{eqnarray}
\left\{\begin{array}{c}
Q \\ L \\ \bU \\ \bD \\ \bE \end{array}\right\}
& \rightarrow & -\;
\left\{ \begin{array}{c}
Q \\ L \\ \bU \\ \bD \\ \bE \end{array}\right\} \nonumber \\
\left\{\begin{array}{c} H_1 \\ H_2 \end{array}\right\}
& \rightarrow & \quad\;
\left\{ \begin{array}{c}
H_1 \\ H_2 \end{array}\right\}
\end{eqnarray}
It is a trivial exercise to show that $R$-parity eliminates the
terms in brackets in the superpotential (\ref{big super}).

\begin{figure}[t]
\hspace*{0.5truein}
\psfig{figure=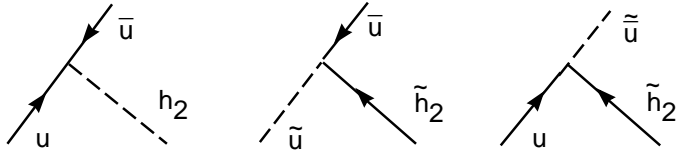,height=1.0in}
\caption{Some of the vertices that arise from the superpotential.
These vertices are all proportional to the up Yukawa coupling.
Because of $R$-parity, the number of supersymmetric particles
in each vertex equals zero, modulo two.}
\label{P vertices}
\end{figure}

In terms of component fields, $R$-parity leaves invariant
the fields of the usual standard model, and flips the sign of
their supersymmetric partners.   Therefore it implies
that supersymmetric particles are pair produced, and that the
lightest supersymmetric particle cannot decay.  (See
Fig.~\ref{P vertices}.)

The superpotential for the MSSM with $R$-parity takes the following
simple form
\begin{equation}
P \;= \;\mu\, H_1 H_2 \; + \; \lambda_U \, Q \bU H_2
\;+\; \lambda_D \, Q \bD H_1 \; + \; \lambda_E \, L \bE H_1\;.
\label{MSSM P}
\end{equation}
The superpotential determines the scalar potential
\begin{equation}
\V\ =\ {1\over 2}\, g_1^2\, D^2\ +\ {1\over 2}\, g_2^2 \,D^{(i) 2}\ 
+\ {1\over 2}\, g_3^2\, D^{(a) 2}\ +\ |\,P_i\,|^2\ ,
\label{potential}
\end{equation}
where the functions $D$ and the superpotential $P$ are specified
above.

As in any field theory, once we have the potential we must look
for its minimum.  Our hope is to find a minimum which preserves
SU(3) $\times$ U(1).  Therefore we shall set $\langle
\tilde q \rangle = \langle \tilde{\bar{u}} \rangle  =  \langle
\tilde{\bar{d}} \rangle = 0$, and consider the following
piece of the full potential,
\begin{eqnarray}
\V & = & +\ {1\over8}\,g^2_1\,\bigg[\,h_1^\dagger h_1
\ -\ h_2^\dagger h_2 \ +\ \tilde{l}^\dagger \tilde{l}
\ -\ 2 \tilde{\bar{e}}^* \tilde{\bar{e}}\,\bigg]^2 \nonumber \\[2mm]
&& +\ {1\over2}\,g^2_2\,\bigg[\, h_1^\dagger T^{(i)} h_1\ +\ 
h_2^\dagger T^{(i)} h_2 \ +\ \tilde{l}^\dagger T^{(i)} \tilde{l} \,
\bigg]^2 \nonumber \\[2mm]
&&+ \ 
\big|\,\lambda_{Eij}\,\tilde{l^i} \tilde{\bar{e}}^j\ +\ \mu\,h_2\,\big|^2
\ +\ \mu^2\,\big|\, h_1\,\big|^2 \nonumber \\[4mm]
&& +\ \big|\,\lambda_{Eij}\,\tilde{\bar{e}}^j h_1 \,\big|^2
\ +\ \big|\,\lambda_{Eij}\, \tilde{l}^i h_1 \,\big|^2 \ +\ ....
\label{MSSM potential}
\end{eqnarray}

It is a straightforward exercise to compute the minimum of this
potential.  At the minimum, one finds that electromagnetism is
not broken,
\begin{equation}
\langle \tilde l \rangle \; = \; \langle \tilde {\bar{e}} \rangle
\; = \; 0\;,
\label{l no}
\end{equation}
which is very good news.  However, one also finds
that electroweak symmetry is not broken,
\begin{equation}
\langle h_1 \rangle \; = \; \langle h_2 \rangle \; = \; 0\ ,
\label{h 1 no}
\end{equation}
which is not.  Equations (\ref{l no}) and (\ref{h 1 no}) imply
that the simplest version of the MSSM does not work.  It
stabilizes $\mu$ against radiative corrections, but it does not
break electroweak symmetry.  Furthermore, all masses are zero,
except for the Higgs supermultiplet, which has a common mass $\mu$.

\mysection{Supersymmetry Breaking}

In the previous section we have seen that the simplest version of
the MSSM leads to a theory in which gauge symmetry is not broken.
In fact, supersymmetry is not broken either.  
This is not acceptable because unbroken supersymmetry requires
the observed particles and their supersymmetric partners to have
the same mass.

In this section we will discuss the spontaneous breaking of
supersymmetry.  We will not discuss explicit supersymmetry breaking
because supersymmetry is a spacetime symmetry, and explicit breaking
leads to inconsistencies when supersymmetry is coupled to supergravity.

\begin{figure}[t]
\hspace*{1.0truein}
\psfig{figure=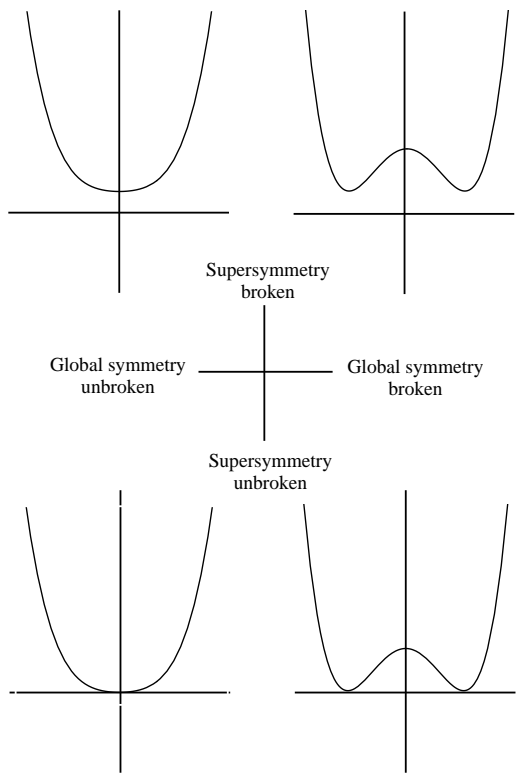,height=4.0in}
\caption{The vacuum energy is the order parameter for
spontaneous supersymmetry\newline breaking.}
\label{susybreak}
\end{figure}

The vacuum energy is the order parameter for spontaneous supersymmetry
breaking.  This can be seen by taking the trace of the supersymmetry
algebra,
\begin{equation}
\{ \, Q_\alpha, \; \bQ_\dalpha \, \} \; = \; 2 \sigma^m_{\alpha \dalpha}
P_m\;.
\end{equation}
One finds
\begin{equation}
\frac{1}{4} \bigg( \, Q_1 \bQ_1 \; + \; \bQ_1 Q_1 \; + \; Q_2
\bQ_2 \;+\;\bQ_2 Q_2 \bigg) \; = \; H\;,
\end{equation}
where $H$ is the Hamiltonian.  The operator on the left-hand side is
positive semidefinite.  Therefore the supercharges annihilate the
vacuum
\begin{equation}
Q \; | \, 0 \, \rangle \; = \; 0
\end{equation}
if and only if the Hamiltonian does as well,
\begin{equation}
H \; | \, 0 \, \rangle \; = \; 0\;,
\end{equation}
provided the Hilbert space has positive norm.
In other words, supersymmetry is unbroken if and only if \cite{mybook}
\begin{equation}
\langle \, \V \, \rangle \; = \; 0\;.
\end{equation}
The situation is summarized in Fig.~\ref{susybreak}.  

For the case at hand, the various contributions to the scalar
potential are of the following form,
\begin{equation}
\V\ =\ {1\over 2}\, g_1^2\, D^2\ +\ {1\over 2}\, g_2^2 \,D^{(i) 2}\ 
+\ {1\over 2}\, g_3^2\, D^{(a) 2}\ +\ |\,P_i\,|^2\ ,
\end{equation}
If $\langle\, P_i\, \rangle \neq 0$ (for some $i$), $\langle\, \V\, \rangle
> 0$ and supersymmetry is spontaneously broken \cite{oraif}.  On the other
hand, if $\langle\,  P_i\, \rangle = 0$, a vacuum can always be found where
$\langle D^{(a)} \rangle = \langle D^{(i)} \rangle = \langle D \rangle
= 0$, so $\langle\, \V\, \rangle = 0$ and supersymmetry is
preserved.\footnote{We ignore the possibility of a Fayet-Iliopoulos
term in the potential \cite{fi}.  Such a term does not change
our conclusions about spontaneous supersymmetry breaking in
the MSSM \cite{wise}.}  Therefore the signal for spontaneous
supersymmetry breaking is that $\langle\, P_i\, \rangle \neq 0$ for
some $i$.

For the case of the MSSM, we previously found a minimum of the
potential at
\begin{equation}
\langle \tilde q \rangle \; = \; \langle \tilde l \rangle \; = \;
\langle \tilde {\bar{u}} \rangle \; = \; \langle \tilde {\bar{d}}
\rangle\; = \; \langle \tilde {\bar{e}} \rangle \; = \; 0
\label{color vevs}
\end{equation}
and
\begin{equation}
\langle h_1 \rangle \; = \; \langle h_2 \rangle \; = \; 0\;.
\label{Higgs vevs}
\end{equation}
Substituting these vevs into the potential, we find that the vacuum
energy is zero, $\langle \, \V \, \rangle = 0,$
which implies that supersymmetry is preserved.

Thus we have seen that the simplest version of the MSSM preserves
supersymmetry {\it and} electroweak symmetry.  Both must be broken.  One
way to do this is to clutter up the theory by adding more fields, which we
reject out of hand.  A second, more appealing approach can be found by
relaxing one of the assumptions that underlie the MSSM.

In the first lecture we motivated weak-scale supersymmetry in terms of the
hierarchy problem.  We presented the MSSM as a fundamental theory in which
the light Higgs mass was protected from destabilizing divergences.  In
what follows, we will keep this motivation, but discard the notion that
the MSSM is a fundamental theory.  Instead, we will view the MSSM as an
{\it effective theory} valid below a scale $M$.

In practical terms this means that the MSSM no longer needs to be
renormalizable.  Indeed, it should contain an infinite tower of
higher-dimensional operators suppressed by the scale $M$.  The full
effective theory is described by an action of the form
\begin{eqnarray}
S & = & \int  d^4x d^4\theta \; K(\Phi^+ e^{gV}, \Phi) \nonumber \\
&& + \; \left[ \, \int d^4x d^2\theta \; \bigg(\,
{1\over4}\,H(\Phi)\, WW\ +\ P(\Phi)\,\bigg) \; + \; 
{\rm h.c.}\, \right]
\end{eqnarray}
where $K(\Phi^+,\Phi)$ is a real function known as the K\"ahler potential
\cite{gensusy}, $H(\Phi)$ is an analytic gauge potential, and $P(\Phi)$ is
the analytic superpotential, each with an expansion in powers of $1/M$:
\begin{eqnarray}
K(\Phi^+,\Phi) & = & \Phi^+\Phi \; + \; \Phi^+ \Phi \,
\left({\Phi + \Phi^+\over M}\right) \; + \; ... \nonumber \\
H(\Phi) & = & 1 \; + \; {1\over M} \, \Phi \; + \; ... \nonumber \\[2mm]
P(\Phi) & = & \frac{1}{2} \, \mu \, \Phi^2 \; +\; \frac{1}{3} \, \lambda 
\,\Phi^3 \; + \; {1\over M} \, \Phi^4 \; + \; ....
\end{eqnarray}
The K\"ahler potential contains generalized kinetic terms, while the
superpotential contains generalized Yukawa couplings.  (In these
expressions, we have not written coefficients of order one in front of
the nonrenormalizable terms.)

For our purposes we do not need to know much about the theory at the scale
$M$.  All we need to assume is that it preserves SU(3) $\times$ SU(2)
$\times$ U(1) and that it breaks supersymmetry at a scale $M_S$.  These two
facts imply that there is a chiral superfield $U$ whose $\theta\theta$
component has a vev of order $M_S^2$,
\begin{equation}
U \; = \; \theta \theta \; M_S^2\ .
\end{equation}
The field $U$ is a spurion whose sole role is to communicate supersymmetry
breaking to the fields of the MSSM.  It contributes to the Lagrangian through 
nonrenormalizable terms suppressed by $1/M$, such as
\begin{eqnarray}
&&{1 \over M^2} \int d^4\theta \; \Phi^+ \Phi \; U^+ U
\nonumber \\[2mm]
&&{1 \over M}  \int d^2\theta \; U\, W^\alpha W_\alpha  \nonumber \\[2mm]
&&{1 \over M}  \int d^2\theta \; \left[ \, \frac{1}{2}\, m \, \Phi^2 \; + \; 
\frac{1}{3}\, g \, \Phi^3
\, \right] \; U\ .
\end{eqnarray}
For the case of the MSSM, these terms introduce a host of new
parameters \cite{susysu5}, \cite{soft break}:
\begin{itemize}
\item
5 independent $3 \times 3$ mass matrices for the squarks and sleptons, 
$$M^2_{0i}{}^j\, A^*_j A^i\ ,$$
as well as two independent masses for the Higgs scalars;
\item
3 independent gaugino masses, 
$$M_{1/2}^a\, \lambda \lambda\ ,$$
for the three factors of the standard model gauge group;
\item
One analytic mass for the two Higgs doublets
$$ \mu B \, h_1 h_2\; ;$$
\item
27 analytic trilinear couplings for the scalar fields, 
$$A_{ijk} A^i A^j A^k\ ,$$
where $A_{ijk} = 0$ unless the coupling is allowed by gauge invariance.
\end{itemize}
For simplicity, we take the soft parameters to be real.
These terms break supersymmetry explicitly in the low energy effective
Lagrangian.  Clearly $M_{0i}{}^j \simeq M_{1/2}^a \simeq \mu B \simeq A_{ijk}
\simeq M_W$ for the hierarchy to be safe from destabilizing divergences.

The soft symmetry breaking operators solve several
of the problems associated with the simplest version of the MSSM.
For example, they lift the masses of the supersymmetric particles
out of reach of present experiments.  They also change the potential
to permit electroweak symmetry breaking,
\begin{eqnarray}
&\langle h_1 \rangle \; = \; v_1 \nonumber \\
&\langle h_2 \rangle \; = \; v_2
\end{eqnarray}
where $v_1, v_2 \neq 0$.

However, the soft supersymmetry breakings introduce their own set of problems.  
They enlarge the parameter space to include over 50 new parameters, so
the MSSM is no longer quite so minimal.  More importantly, the soft operators
can induce rare processes such as flavor-changing neutral currents
\cite{susysu5}, \cite{fcnc}.
The operators must be carefully constrained.

\begin{figure}[t]
\hspace*{1.1truein}
\psfig{figure=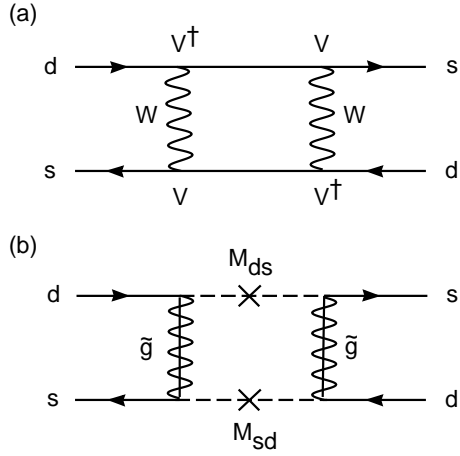,height=2.5in}
\caption{Diagrams that contribute to $K-\bar K$ mixing.  (a)  The
standard-model contributions are suppressed by the GIM mechanism
because $V V^\dagger = 1$.  (b)  The squark mass matrices give rise
to supersymmetric contributions to the mixing.}
\label{kkbar}
\end{figure}

To illustrate the problem, let us examine the canonical example of $K-
\bar{K}$ mixing.   We will work in a supersymmetric basis, in which the quark
mass matrices are diagonal.  Then the usual contributions to $K-\bar{K}$
mixing are suppressed by the GIM mechanism, as shown in Fig.~\ref{kkbar}(a).

In supersymmetric theories there are additional diagrams which contribute
to $K-\bar{K}$ mixing.  A gluino contribution is shown in Fig.~\ref{kkbar}(b).
In this diagram the flavor changing neutral current (FCNC) is induced by the
squark mass matrix.  From the diagram one can see that the FCNC vanishes if
the LL and RR entries of the squark mass matrices are proportional to
the identity, and the LR entries are proportional to the 
Yukawa matrix, $\lambda_D$.  Then the rotations which diagonalize the
quark mass matrix,
\begin{eqnarray}
u & \rightarrow & V_u \, u \nonumber \\
d & \rightarrow & V_d \, d \nonumber \\
\bar{u} & \rightarrow & V_{\bar{u}} \, \bar{u} \nonumber \\
\bar{d} & \rightarrow & V_{\bar{d}} \, \bar{d}\ ,
\label{squark rotns}
\end{eqnarray}
also eliminate all terms which connect $\tilde{s}$ to $\tilde{d}$.

A second way to suppress the FCNC is to take the soft LL and RR mass
matrices to be proportional to the Yukawa matrices themselves \cite{alignn}.  
Then the LL and RR terms are of the form
\begin{eqnarray}
& \tilde{q}^\dagger \, ( \lambda^*_U \lambda^T_U \; + \lambda^*_D\;
\lambda^T_D) \; \tilde{q} & \nonumber \\[1mm]
& \tilde{\bar{u}}^\dagger \, ( \lambda^\dagger_U \lambda_U ) \,
\tilde{\bar{u}} & \nonumber \\[1mm]
& \tilde{\bar{d}}^\dagger \, ( \lambda^\dagger_D \lambda_D ) \,
\tilde{\bar{d}} \ , &
\label{align}
\end{eqnarray}
where the Yukawa couplings are matrices in flavor space.  These
Yukawa terms give rise to the following squark mass matrices,
\begin{eqnarray}
& \tilde{u}^\dagger \, (M_u^2 \; +\; V M^2_d \; V^\dagger ) \, 
\tilde{u} \nonumber \\[1mm]
& \tilde{d}^\dagger \, ( M_d^2 \; +\; V^\dagger M^2_u \; ) \, 
\tilde{d} \nonumber \\[1mm]
& \tilde{\bar{u}}^\dagger \, ( M_u^2 ) \, 
\tilde{\bar{u}}\;\; \nonumber \\[1mm]
& \tilde{\bar{d}}^\dagger \, ( M_d^2 ) \, 
\tilde{\bar{d}}\ ,
\end{eqnarray}
where $M_u$ and $M_d$ are the diagonalized up- and down-type quark mass
matrices, and $V \equiv V^\dagger_u V_d$ is the usual CKM matrix.  For
soft masses of the form (\ref{align}), the flavor changing neutral currents
are suppressed by a supersymmetric generalization of the usual GIM mechanism.

\mysection{Naturalness, Revisited}

In the previous section we have seen that supersymmetry and gauge symmetry
can be broken by operators which arise if the MSSM is an effective theory,
valid below a scale $M$.  In this section we will revisit the hierarchy
problem to make sure that the Higgs stays light even though
another scale has been introduced into the theory \cite{dest 1}, \cite{dest 2}.    
We will see whether radiative corrections still respect the electroweak hierarchy.

The subject of supersymmetric radiative corrections is rather technical,
involving perturbation theory in superspace (or, involving subtle questions
of regularization in components) \cite{mybook}.  The end result is that
the K\"ahler potential can receive perturbative radiative corrections.
\begin{equation}
\int d^4\theta \; K \ \rightarrow\ \int d^4\theta \; K \; + \;
\int d^4\theta \; \delta K\;.
\end{equation}
The superpotential, however, cannot
\begin{equation}
\int d^2 \theta \; P \ \rightarrow\ \int d^2 \theta \; P\ .
\end{equation}
The supersymmetric nonrenormalization theorem states that the superpotential
receives no corrections at all -- not finite, not infinite -- to any order in
perturbation theory.\footnote{In some cases, the superpotential can receive
nonperturbative corrections.  See the lectures of Seiberg \cite{seiberg} for
more on this subject.}

The standard proof of the nonrenormalization theorem requires superfield
perturbation theory, which is too technical for these lectures.  Instead,
let us prove the theorem in a manner discussed by Seiberg \cite{nati nonren}.  
Consider the Lagrangian
\begin{equation}
\int d^4\theta \; \Phi^+_i \Phi^i\ +\ \;\Big[ \int d^2\theta \,
\Big(\, {1\over2} \, m_{ij} \, \Phi^i\Phi^j
\; + \; {1\over3} \, \lambda_{ijk} \, \Phi^i \Phi^j\Phi^k\,\Big) \;+
\;{\rm h.c.}\Big]  \;.
\label{un action}
\end{equation}
In what follows, we will think of $m_{ij}$ and $\lambda_{ijk}$ as the
vev's of classical background superfields.  In other words, we will
take their kinetic energies to be
\begin{equation}
\lim_{\Lambda\rightarrow\infty} \; \Lambda^2 \; \int d^4\theta\;
(m^{ij+} m_{ij} \; + \; \lambda^{ijk+} \lambda_{ijk})\;,
\end{equation}
in which case the fields have dimension zero and do not propagate.

The action (\ref{un action}) is manifestly invariant under a global
U(N) symmetry.  It is also invariant under a continuous 
$R$-symmetry, with $R$-charges assigned as follows:
\begin{eqnarray}
\theta & \rightarrow & e^{-i\alpha} \; \theta \nonumber \\
\Phi^i & \rightarrow & e^{i \alpha} \; \Phi^i \nonumber \\
m_{ij} & \rightarrow & m_{ij} \nonumber \\
\lambda_{ijk} & \rightarrow & e^{-i \alpha}\; \lambda_{ijk}\;.
\end{eqnarray}
The U(N) $\times$ U(1)$_R$ symmetry plays a major role in constraining
the quantum corrections.

As a first step towards proving the theorem, we consider the
renormalization of the $\Phi^3$ term in the superpotential.  At one loop,
the correction cannot involve $\lambda^+$ or $m^+$ because the
superpotential must be analytic.  Therefore the only U(N) invariant
is of the form
\begin{equation}
\lambda_{...} \lambda_{...} \lambda_{...}\;m^{-1..} m^{-1..}
\;\Phi^.  \Phi^.  \Phi^.\;,
\end{equation}
where the dots denote U(N) indices contracted in different ways.
The problem with this term is that it violates $R$-symmetry.  More
insertions of $\lambda$ makes this even worse, so there can be no
renormalization of the $\Phi^3$ coupling.  (Nonperturbative corrections
of the form $\exp(-m/\lambda\Phi)$ are not permitted because they are
singular at weak coupling for negative $\Phi$.)

\begin{figure}[t]
\hspace*{1.1truein}
\psfig{figure=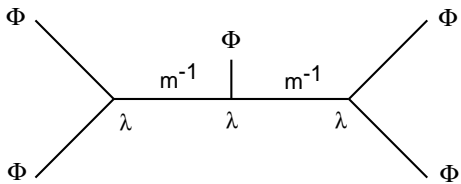,height=1.25in}
\caption{This $\Phi^5$ diagram is not 1PI and does not contribute
to the renormalized\newline superpotential.}
\label{tree}
\end{figure}

Now let us consider a higher-dimensional operator, such as a possible
$\Phi^5$ coupling.  A contribution of the form
\begin{equation}
\lambda_{...} \lambda_{...} \lambda_{...}\;m^{-1..} m^{-1..}
\;\Phi^.  \Phi^.  \Phi^.  \Phi^.  \Phi^.\;
\label{phi5}
\end{equation}
is U(N) $\times$ U(1)$_R$ invariant.  However, this term corresponds
to the diagram of Fig.~\ref{tree}.  This diagram is not 1PI, so it does
not correspond to a term in the renormalized superpotential.

These arguments can be readily extended to all other operators.  For the
case at hand, the superpotential is not renormalized, either perturbatively
or\break\newpage nonperturbatively, because of
\begin{enumerate}
\item
analyticity,
\item
global U(N) symmetry,
\item
global U(1)$_R$ symmetry, and
\item
a smooth weak-coupling limit.
\end{enumerate}

Let us now apply the nonrenormalization theorem to the study of
naturalness in supersymmetric theories.  The theorem tells us that
all potentially destabilizing renormalizations are corrections to
the K\"ahler potential. To classify the dangerous diagrams, we need
to determine the superspace degree of divergence.

Superspace power counting is not hard to derive.  A diagram with $E_\Phi$
external chiral superfields has the following cutoff dependence,
\begin{equation}
\Lambda^D \; \int  d^4 \theta \; \Phi^+\;... \; \Phi\;,
\end{equation}
where $D \leq 2 - E_\Phi + \sum d V_d$, and $V_d$ denotes the number
of nonrenormalizable operators suppressed by $(1/M)^d$.  If we include
the factors of $1/M$, we see that the divergence associated with a
given diagram goes like
\begin{equation}
\Lambda^D\;\prod_d \, \left( {1\over M} \right)^{d V_d} \;\lsim
\;M^{2 - E_\Phi}\ ,
\end{equation}
for $ \Lambda \lsim M$.  Superspace power counting indicates that the only
dangerous diagrams are tadpoles, with $E_\Phi = 1$.

To see why tadpoles are dangerous, let us consider a specific example in which
we restrict our attention to a single ``Higgs" superfield, $H$.  We will let
$N$ be a gauge- and global-symmetry singlet chiral superfield which couples
directly to the Higgs.  Therefore we will take the superpotential, $P$, to be
\begin{equation}
P \;=\; {1 \over 2}\,\mu\, H^2\;+\;{1 \over 2}\, m\, N^2 \;+
\;\lambda\, N H^2\;+\;...,
\label{PN}
\end{equation}
where we fix the Higgs mass $M_h = \mu \simeq m \simeq M_W$.  
(A discrete $Z_2$ symmetry replaces the gauge symmetry of the standard
model.  We assume that $Z_2$ is not broken for scales larger than
$M_W$.)  The hierarchy is destabilized if radiative corrections
lift $M_h \gg M_W$.

Now let us suppose that our theory is a low energy effective theory,
coupled by nonrenormalizable operators to the spurion $U$.  In this
case, the K\"ahler potential becomes
\begin{equation}
K \;=\;\bigg[\,N^+N\;+\;H^+H\;+\;H^+H\,\bigg({N + N^+ \over M}\bigg)\,\bigg]
\bigg\{ \, 1\;+\;\bigg({U + U^+ \over M}\bigg)\;+\;...\,\bigg\}
\label{PC}
\end{equation}
where we have neglected coefficients of order one.  Typically, the
fields $H$ and $N$ have weak-scale vevs,
\begin{eqnarray}
\langle H \rangle & \lsim &  M_W \;+\;\theta \theta M_W^2 \nonumber \\
\langle N \rangle & \lsim &  M_W \;+\;\theta \theta M_W^2\;,
\label{hvev}
\end{eqnarray}
while
\begin{equation}
\langle U \rangle\;\simeq\;\theta \theta M_S^2\;.
\label{cvev}
\end{equation}
The vevs (\ref{hvev}) and (\ref{cvev}) preserve hierarchy, as can be
seen by substituting into (\ref{PN}) and (\ref{PC}).  They induce a
supersymmetry-breaking mass of order $M_W$ for the scalar component
of the Higgs superfield.

\begin{figure}[t]
\hspace*{1.1truein}
\psfig{figure=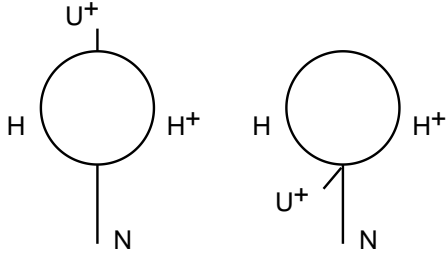,height=1.5in}
\caption{Potentially dangerous tadpole diagrams.}
\label{tadpole}
\end{figure}

At one loop, these vevs can shift.  In the above example, there are two
potentially dangerous superspace diagrams, as shown in Fig.~\ref{tadpole}.
Each $U^+$ insertion induces a quadratic divergence
\begin{equation}
\delta S \ \simeq \ {\Lambda^2 \over M^2}\,\int d^4x d^4\theta\; U^+ N\;+
\;....
\end{equation}
Taking the cutoff $\Lambda \simeq M$, we find
\begin{eqnarray}
\delta S & \simeq & \int d^4x d^4\theta\;U^+ N\;+\;...\nonumber\\
& \simeq & M^2_S \int d^4x d^2\theta \;N\;+\;....
\end{eqnarray}
This term induces a vev of order $M^2_S$ for $F_N$, which in turn gives
rise to masses of order $M_S$ for the scalar fields $n$ and $h$.  The
hierarchy is, in fact, destabilized.

This example illustrates that the hierarchy can be destabilized
when a second scale is introduced into the theory.  However, the
destabilization requires a gauge- and global-symmetry singlet, so the
MSSM is safe.  The next-to-minimal standard model is not necessarily
safe because it contains a singlet superfield, $N$.

\begin{figure}[t]
\hspace*{1.1truein}
\psfig{figure=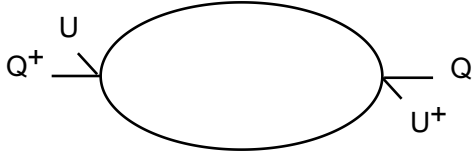,height=0.9in}
\caption{A quadratically divergent renormalization of the soft
squark mass.}
\label{squarkren}
\end{figure}

Even for the MSSM, however, the quadratically divergent
radiative corrections carry an important lesson: the soft
supersymmetry-breaking parameters cannot be calculated in terms
of the low-energy effective field theory.  They depend sensitively
on physics at the scale $M$.  This can be seen by considering the
following terms in the K\"ahler potential,
\begin{equation}
{1\over M}\,\bigg[ \,
\lambda_U\, Q \bU H^+_1\;+\;\lambda_D\, Q \bD H^+_2 \,\bigg]
\ \bigg\{\,1\;+\;\bigg({U + U^+ \over M}\bigg)\,\bigg\}\;.
\end{equation}
These terms give rise to quadratically divergent diagrams such as
those in Fig.~\ref{squarkren}.  When reduced to components, they
give rise to additive renormalizations of the squark masses, such as
\begin{equation}
M^4_S\,\left({\Lambda^2 \over M^4}\right)\,
\tilde{\bar{u}}^\dagger\, ( \lambda^\dagger_U \lambda_U )
\,\tilde{\bar{u}}\;.
\end{equation}

This operator has the same flavor structure as in eq.~(\ref{align}).
For $\Lambda \lsim M$, it does not destroy the hierarchy.  However,
the quadratic divergence tells us that the coefficients of the soft
supersymmetry breaking operators cannot be calculated in terms of the
low energy effective theory.  They depend on physics at the scale $M$,
and must be fixed by matching conditions at that scale \cite{matching}.

\mysection{Electroweak Symmetry Breaking}

In the previous section we have seen that the MSSM with arbitrary
soft supersymmetry breaking contains over 50 new parameters.  Indeed,
it may well be that nature adjusts each of them independently to
describe the physical world.  However, as a first step towards
understanding the  phenomenology of supersymmetric models, it makes
sense to shrink the parameter space to a more manageable size.

\begin{figure}[t]
\hspace*{0.8truein}
\psfig{figure=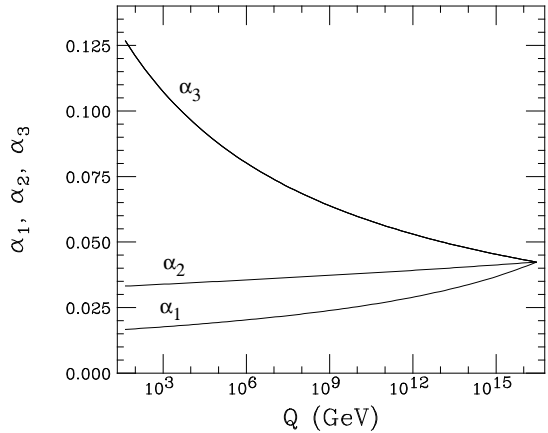,height=2.5in}
\caption{In supersymmetric theories, the running gauge couplings unify
at the scale $M_{\rm GUT} \simeq 10^{16}$ GeV.}
\label{unify}
\end{figure}

Since the soft symmetry breakings originate at the scale $M$,
restrictions on the parameters amount to assumptions about
physics at that scale \cite{susyradbreak}, \cite{sugramodels}.
Therefore in what follows we will be
motivated by the fact that -- in supersymmetric theories -- the
running gauge couplings unify at a scale $M_{\rm GUT}
\simeq 10^{16}$ GeV \cite{susysu5},
\cite{susysu5b}, \cite{predict}, \cite{postdict},
as shown in Fig.~\ref{unify}.  In light of this, 
it is reasonable to assume that the soft parameters unify as
well,  in which case they are completely specified by four
parameters at the scale $M_{\rm GUT}$,
\begin{enumerate}
\item One common scalar mass, $M_0$;
\item One common gaugino mass, $M_{1/2}$;
\item One analytic Higgs mass, $B\mu$;
\item One trilinear coupling, $A_0\, \lambda_F$;
\end{enumerate}
where $A_0$ is the soft parameter and $\lambda_F$ is the appropriate
Yukawa coupling from the superpotential.

Of course, experimental physics is done at the weak scale, so
these parameters must be evolved to $M_W$ using the renormalization
group equations \cite{radbreak}.  This is fortunate because -- if the
scalar masses, $M_0$, were degenerate at the weak scale -- either
no gauge symmetries would be broken, or all would be broken.

Thus, at the weak scale, the effective potential is of the form
\begin{eqnarray}
\V & =& \
{1\over 2}\, g_1^2\, D^2\;+\;{1\over 2}\, g_2^2\, D^{(i) 2}\;
+\;{1\over 2}\, g_3^2\, D^{(a) 2} \;+\;|\,P_i\,|^2\;\nonumber \\[3mm]
&& +\;M^2_{Q}\,\tilde{q}^{\dagger} \tilde{q}\;+
\;M^2_{\bar{U}}\,\tilde{\bar{u}}^{\dagger }\tilde{\bar{u}}\;+
\;M^2_{\bar{D}}\,\tilde{\bar{d}}^{\dagger }\tilde{\bar{d}}
\nonumber \\[4mm]
&& +\;M^2_{L}\,\tilde{l}^{\dagger} \tilde{l}\;+
\;M^2_{\bar{E}}\,\tilde{\bar{e}}^{\dagger }\tilde{\bar{e}}\;+
\;M^2_1 \, h_1^\dagger   h_1\;+\;
M^2_2\,h_2^\dagger h_2 \nonumber \\[4mm]
&& +\;\bigg\{\,\bigg[ A^U_{ij}\, \tilde{q}^i
\tilde{\bar{u}}^{\,j} h_2
\;+ \;A^D_{ij}\,\tilde{q}^i
\tilde{\bar{d}}^{\,j} h_1\;+\;
A^E_{ij}\,\tilde{l}^i \tilde{\bar{e}}^{\,j} h_1 \nonumber\\
&& + \;\mu\,B\,h_1 h_2\, \bigg]\;+\;{\rm h.c.}\;\bigg\}\ ,
\label{weakpot}
\end{eqnarray}
where, at the unification scale, we impose the boundary condition
\begin{eqnarray}
M^2_Q \ =\  M^2_{\bar{U}} \ =\  M^2_{\bar{D}} \ =\  M^2_L &=&
M^2_{\bar{E}}\ =\ M^2_1 \ =\  M^2_2 \ \equiv\  M^2_0 \nonumber\\[2mm]
A^U_{ij}\ \equiv\ \lambda_{Uij}\,A_0 \qquad A^D_{ij} &\equiv&
\lambda_{Dij}\,A_0\ \qquad  A^E_{ij}\ \equiv\ \lambda_{Eij}\,A_0
\nonumber\\[2mm]
B\mu &\equiv & B\mu \ .
\label{massparms}
\end{eqnarray}

The full computation of the effective potential is beyond the scope
of  these lectures.  To grasp the idea, however, we will focus on the 
most important corrections to the soft scalar masses. 
For this purpose, it is sufficient to consider the effects of the
top Yukawa $\lambda_T$.  (The strong gauge coupling does not
contribute to the running of the squark masses at one loop.)

\begin{figure}[t]
\hspace*{0.5truein}
\psfig{figure=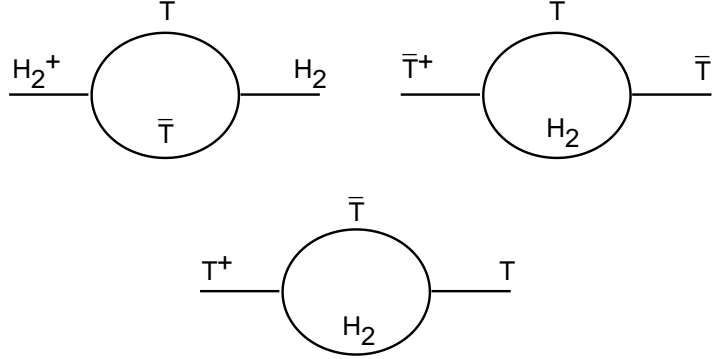,height=2.0in}
\caption{Diagrams that contribute to the running of the soft
squark masses.  Each diagram requires insertions of the spurions
$U$ and $U^+$ on its lines and vertices.}
\label{rge}
\end{figure}

The top Yukawa $\lambda_T$ links the fields $H_2$, $T$ and $\bar{T}$
in the superpotential.  These couplings renormalize the mass parameters
$M^2_2$,  $M^2_T$ and $M^2_{\bar{T}}$ through diagrams like those of
Fig.~\ref{rge}.  The resulting renormalization group equations are
\cite{wise,radbreak}
\begin{equation}
Q\,{d\over dQ}\,
\pmatrix{ M^2_2 \crr M^2_{\bar{T}} \crr M^2_T \crr}
\;=\;{\lambda^2_T\over 8 \pi^2}\,
\pmatrix{ 3 & 3 & 3 \crr 2 & 2 & 2
\crr 1 & 1 & 1 \crr} 
\pmatrix{ M^2_2 \crr M^2_{\bar{T}} \crr M^2_T \crr}
\;+ \;{\lambda^2_T\over 8 \pi^2}\,|\,A^T\,|^2\,\pmatrix{ 3 \crr 2
\crr 1 \crr}\;,
\end{equation}
where the factors of three come from the three colors running around the
loop of Fig.~\ref{rge}.  Likewise, the factors of two come from SU(2).  
(The color coupling does not contribute to the renormalization group
equations at this order.)

\begin{figure}
\vbox{
\centerline{
\psfig{figure=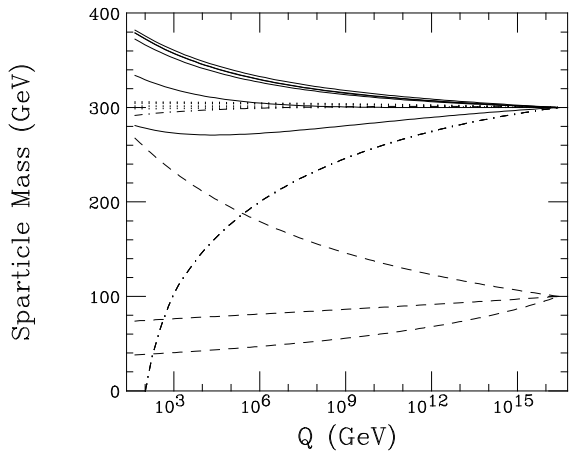,height=2.50in}
}}
\caption{A sample spectrum in the radiative breaking scenario.
Here $M_0 = 300$ GeV and $M_{1/2} = 100$ GeV.  The solid lines
denote squark masses and the dotted lines sleptons.  (The lightest
squark is predominantly $\bar{t}$.)  The dashed lines represent
gaugino masses, while the dot-dashed line marks the mass of the
second Higgs.}
\label{spectrum}
\end{figure}

To analyze this equation, let us forget that $\lambda_T$ runs, and
also ignore the term with $A^T$, for simplicity.  Then the evolution
of the masses is determined by the matrix 
\begin{equation}
\pmatrix{ 3 & 3 & 3 \cr 2 & 2 & 2 \cr
1 & 1 & 1 \cr} \;.
\end{equation}
This matrix has eigenvalues $(0,0,6)$.  At the Planck scale, the 
initial condition on the soft masses can be written in terms of
the eigenvectors,
\begin{eqnarray}
\pmatrix{ M^2_2 \crr M^2_{\bar{T}} \crr M^2_T \crr}& =&
 \;M^2_0\,\pmatrix{ 1 \crr 1\crr 1 \crr} \nonumber \\[5mm]
&=&  \;{1\over2}\,M^2_0\,\left[ \pmatrix{ 0 \crr -1\crr 1 \crr}\right.
\ +\ \pmatrix{ -1 \crr 1\crr 0\crr}\ +
\ \left.\pmatrix{ 3 \crr 2\crr 1\crr}\right]\;.
\end{eqnarray}

The last eigenvector corresponds to the eigenvalue $6$, so it is
damped out during the renormalization from $M$ to $M_W$.  The other 
eigenvectors have eigenvalues zero, so they barely run.  Therefore,
at $M_W$, we expect to find
\begin{equation}
\pmatrix{ M^2_2 \crr M^2_{\bar{T}} \crr M^2_T \crr}\;\simeq
\;{1\over2}\,M^2_0\,\pmatrix{ -1 \crr 0 \crr 1 \crr}\;.
\end{equation}
We see that the renormalization group evolution has flipped the sign
of the $h_2$ mass term.  The large top Yukawa has destabilized the
vacuum: the effective potential breaks SU(2) $\times$ U(1) down 
to the U(1) of electromagnetism!

The effect of the renormalization group evolution on the supersymmetric
mass spectrum is shown in Fig.~\ref{spectrum}, where we plot some
of the running supersymmetric masses between the weak and unification
scales.  Indeed, as expected, the mass (squared) of the second Higgs
is driven negative, and the right-handed top squark is lighter than
the others.

\begin{figure}
\vbox{
\centerline{
\psfig{figure=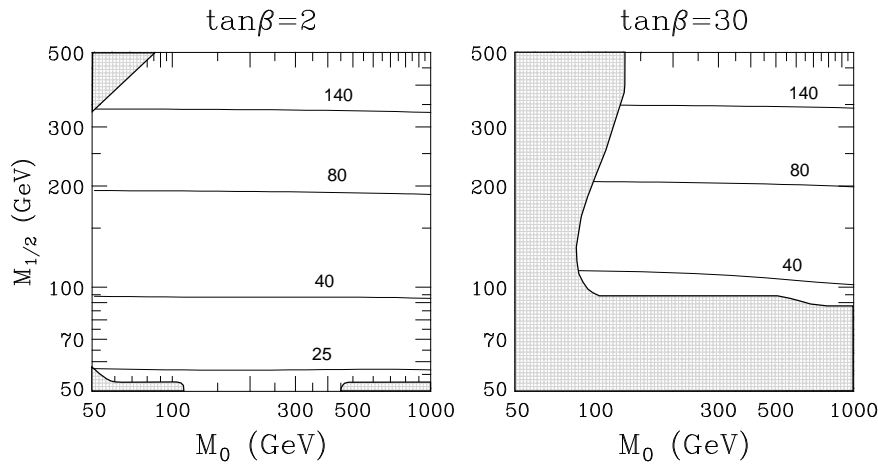,height=2.75in}
}}
\caption{The mass of the lightest supersymmetric particle,
$\chi^0_1$, for $\mu > 0$, $A_0 = 0$, $\alpha_s(M_Z) =
0.12$ and $m_t = 175$ GeV.  The shaded region is forbidden
by experimental and theoretical constraints.  Most of the
supersymmetric parameter space is still open.}
\label{lsp}
\end{figure}

Thus we have seen that in this theory, electroweak symmetry breaking
is driven by a  generalization of the Coleman-Weinberg mechanism
\cite{coleman},
where the large radiative corrections are induced by the top mass.
This mechanism requires $M_t \simeq 175$ GeV.  This remarkable
fact links electroweak symmetry breaking to the presence of a heavy
top!~\footnote{When these models were first proposed in the early
1980's, people thought the top mass would be about 35 GeV, so 
supersymmetry model-builders invented baroque models to make the
top sufficiently light.  If the model-builders had stood their
ground, theorists could have claimed to have {\it predicted} the
mass of the top!}

\mysection{Experimental Expectations}

In what follows, we present expectations for the supersymmetric
spectrum based on this unification scenario.  (For more
details, see the lectures of Tata \cite{tata}.)  Since electroweak
symmetry is broken, we shall trade the parameters $g_1,\;
g_2,\;\mu$ and $B\mu$ for the mass of the $Z$, $M_Z$, the Fermi 
coupling, $G_F$, the fine structure constant, $\alpha_{EM}$, and
the ratio of vevs, $\tan\beta = v_2/v_1$.  We take the strong
coupling,  $\alpha_s$, and the ordinary fermion masses to be given
by their experimental values.

\begin{figure}
\vbox{
\centerline{
\psfig{figure=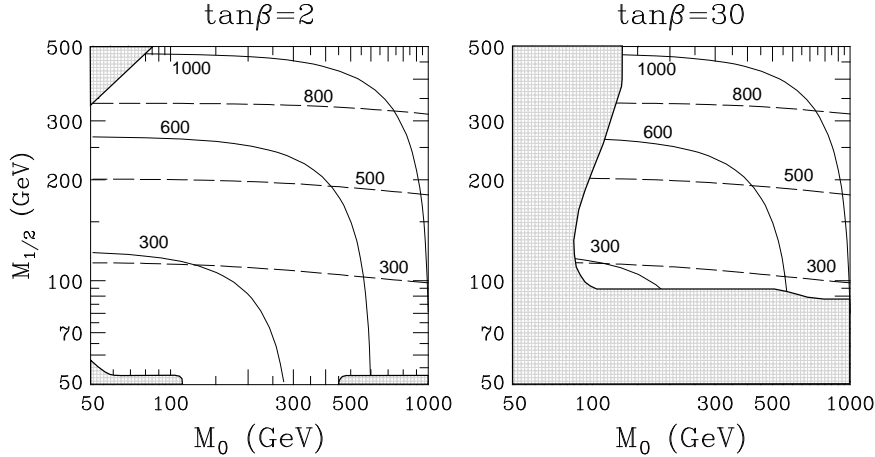,height=2.75in}
}}
\caption{The mass of the up squark (solid line) and the
gluino (dashed line), for $\mu > 0$, $A_0 = 0$, $\alpha_s(M_Z)
= 0.12$ and $M_t = 175$ GeV.  The parameter space
corresponds to squark masses of less than about one
TeV.}
\label{squark}
\end{figure}

In this way we can compute the supersymmetric masses and couplings
in terms of the parameters
\begin{equation}
M_0 \qquad M_{1/2} \qquad A_0 \qquad \tan\beta
\end{equation}
and the sign of $\mu$.  For simplicity, we shall set $A_0 = 0$
and take the supersymmetric Higgs mass parameter $\mu > 0$.

In Fig.~\ref{lsp} we show mass contours for the lightest
superparticle, $\chi^0_1$.  The $\chi^0_1$ is neutral and stable
(because of R-parity).  In the figure, the shaded areas represent
forbidden regions of parameter space, either because of present
experimental limits or because of theoretical constraints such as
the cosmological requirement that the lightest (stable) superparticle
be neutral, or the phenomenological constraint that electroweak
symmetry be broken, but not color.

In Fig.~\ref{squark} we show contours for the (up) squark and
the gluino masses.  (The masses of the up, down, charm and strange
squarks are almost degenerate.)  From the plot we see that the
parameter space covers squark masses up to about 1 TeV.  This is
the range of interest if supersymmetry is to solve the hierarchy
problem. (The rule of thumb is that $M_{\tilde g} \simeq 3 M_{1/2}$
and  $M^2_{\tilde q} \simeq M^2_0 + 4 M^2_{1/2}$.)

\begin{figure}
\vbox{
\centerline{
\psfig{figure=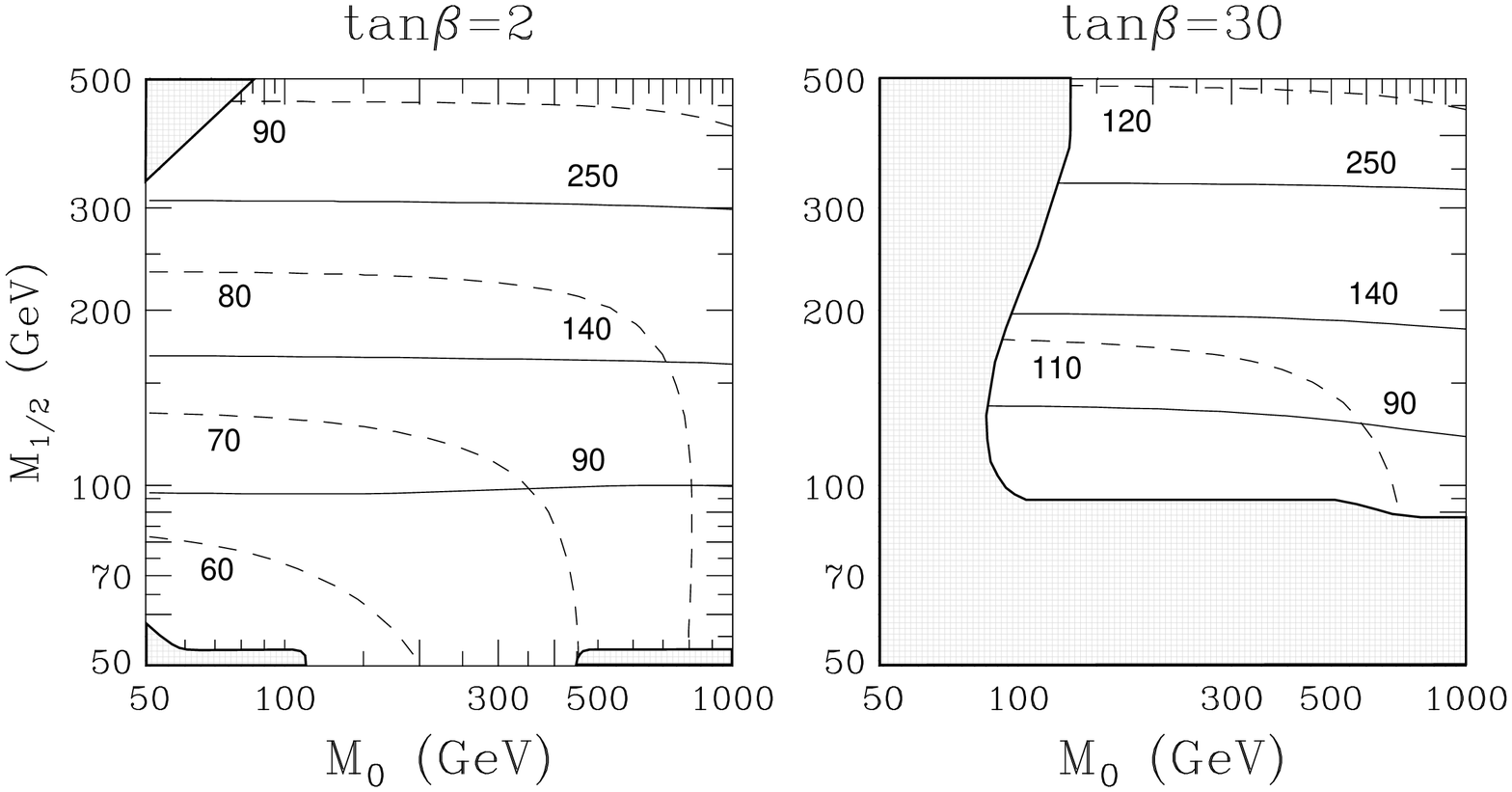,height=2.75in}
}}
\caption{The mass of the lightest chargino, $\chi^\pm_1$,
(solid line) and lightest Higgs, $h$, (dashed line), for $\mu
> 0$, $A_0 = 0$, $\alpha_s(M_Z) = 0.12$ and $M_t = 175$ GeV.
The Higgs mass is less than about 120 GeV over the
parameter space.}
\label{chargino}
\end{figure}

In Fig.~\ref{chargino} we plot contours for the masses of the
lightest Higgs scalar, $h$, and the lightest chargino, $\chi^\pm_1$.  
We see that $M_{\chi^\pm} \simeq  M_{1/2}$, and that 
the maximum Higgs mass is about 120 GeV.  (For completeness,
we note that the slepton masses are approximately $M_L \simeq M_0$.)

\begin{figure}
\vbox{
\centerline{
\psfig{figure=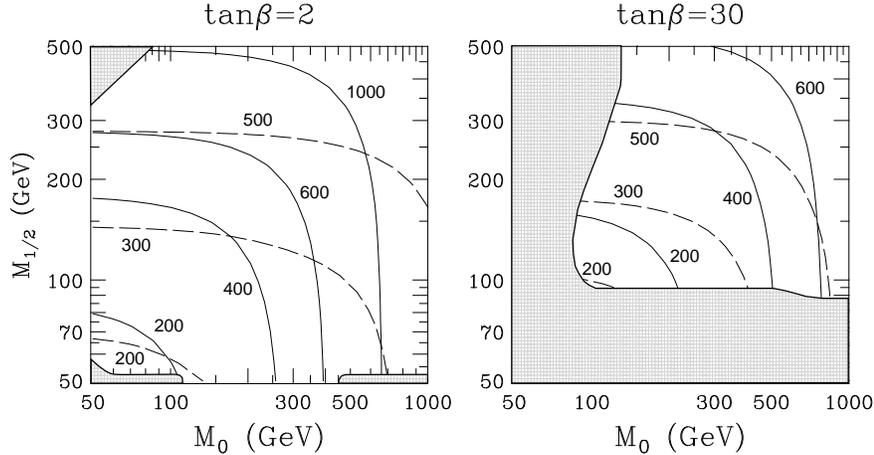,height=2.75in}
}}
\caption{The mass of the charged Higgs, $H^\pm$,
(solid line) and lightest top squark, $\tilde t_1$, (dashed line),
for $\mu > 0$, $A_0 = 0$, $\alpha_s(M_Z) = 0.12$ and $M_t =
175$ GeV.  The decays $t \rightarrow
\tilde t \tilde \chi^0_1$ and $t \rightarrow H^+ b$ are
kinematically forbidden over most of the parameter space.}
\label{stop}
\end{figure}

Finally, in Fig.~\ref{stop} we show contours for the masses of
the lightest top squark, $\tilde t_1$, and charged Higgs,
$H^\pm$.  From the figure we see that the decays $t
\rightarrow \tilde t_1 \tilde \chi^0_1$ and $t \rightarrow H^+
b$ are kinematically forbidden over most of the parameter space.  
(The top squark can be lighter for $A_0 \ne 0$, but a very light stop
requires a fine tuning of the parameters.)

\begin{figure}[t]
\hspace*{0.8truein}
\psfig{figure=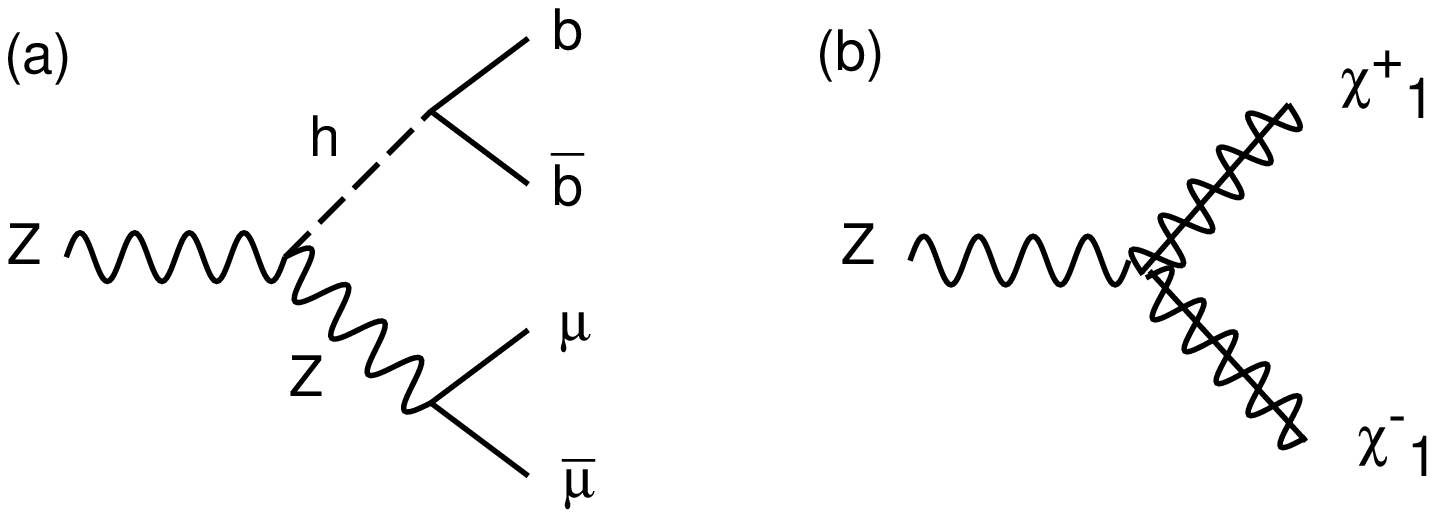,height=1.25in}
\caption{Sample processes contributing to (a) Higgs, and (b)
chargino, discovery\newline at LEP 200.}
\label{lep}
\end{figure}

These figures can be used to illustrate the supersymmetry reach
of a given accelerator.  For example, LEP 200 has a mass reach
of about $\sqrt s - 100$ GeV for a supersymmetric Higgs particle,
and $\sqrt s /2$ for a chargino  \cite{lep200}.  (Sample processes are
illustrated in Fig.~\ref{lep}.)  Therefore
Fig.~\ref{chargino} shows that LEP 200 has an excellent chance of
discovering the lightest supersymmetric Higgs, and a reasonable
possibility of finding the lightest chargino.

\begin{figure}[t]
\hspace*{0.3truein}
\psfig{figure=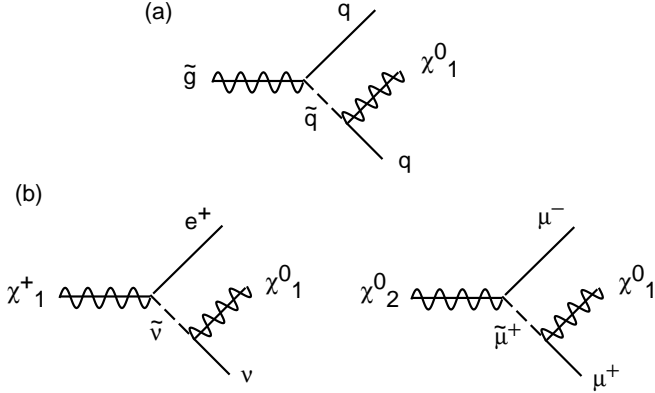,height=2.25in}
\caption{Sample processes contributing to (a) gluino, and (b)
chargino, discovery\newline at the Tevatron.}
\label{fnal}
\end{figure}

The Tevatron's discovery potential is more model-dependent, and
varies considerably with the Tevatron luminosity.  For an integrated
luminosity between 200 pb${}^{-1}$ and 25 fb${}^{-1}$, the gluino
discovery reach is in the range of 300 -- 400 GeV.  Likewise, the
chargino/neutralino reach varies between 150 -- 250 GeV in the
trilepton decay channel, $\chi^+_1 \chi^0_2 \rightarrow \ell^+
\ell^- \ell^{\prime +}$ plus missing energy \cite{tev}.  (Sample
processes are illustrated in Fig.~\ref{fnal}.)  From
Figs.~\ref{squark} and \ref{chargino} we see that an
upgraded Tevatron would begin to cover a significant amount
of the supersymmetric parameter space.

Finally, the LHC has an immense discovery potential.  Assuming 10 fb$^{-1}$
of luminosity, recent studies indicate that the LHC's reach for gluinos
extends significantly past 1 TeV \cite{lhc}.  There are
promising signals in the jets
plus missing energy channel, as well as in channels with leptons and missing
energy.  Clearly, understanding LHC signals and backgrounds is of
enormous importance for supersymmetry.  The great energy of LHC
collisions offers unparalleled opportunities for supersymmetry
discovery.

\mysection{Conclusions}

These lectures presented an introduction to the theory and practice
of weak-scale supersymmetry.  We motivated the subject in terms of
the hierarchy problem, the instability of the Higgs mass to quadratically
divergent radiative corrections.   We found that supersymmetry renders
the Higgs mass natural, and gives rise to a rich new spectroscopy at
the TeV scale.  For every particle of the standard model, supersymmetry
predicts another that has yet to be observed.

Exact supersymmetry implies Bose-Fermi mass degeneracy, so the question 
of supersymmetry breaking is of paramount importance for supersymmetric
theories.  During the course of the lectures we found that the soft
supersymmetry breakings lift the masses of the supersymmetric particles
into a phenomenologically acceptable range.  Soft supersymmetry breaking
suggests that we think of the supersymmetric standard model as an
effective field theory, valid below some scale, $M$.  From this point of
view, supersymmetry breaking occurs at the scale $M$, and gives rise to
soft operators at the scale $M_W$.

With LEP 200, the Fermilab Main Injector and the LHC, prospects look
bright for future experiments.  These accelerators will, for the first
time, begin to probe large regions of the supersymmetric parameter space.  
Ultimately, experiments must say whether supersymmetry is correct.  If
it is, theorists and experimentalists must search for clues to the origin
of supersymmetry breaking -- the central question behind the MSSM.

\section*{Acknowledgments}

It is a pleasure to thank K.T. Mahanthappa and David Soper for organizing
the TASI school, and the TASI students for the many excellent
questions they asked.  I would also like to thank my collaborators,
especially
Sasha Galperin,
Konstantin Matchev,
Sam Osofsky,
Damien Pierce,
Erich Poppitz,
Lisa Randall and
Renjie Zhang,
for sharing with me their many insights about supersymmetry. 
Finally, I would like to thank Edwin Lo, Tom Mehen and Nicolao
Fornengo for  their close reading of this manuscript.  This work
was supported by the U.S. National Science Foundation under grant
NSF-PHY-9404057.

\section*{Appendix}
\setcounter{section}{1}
\renewcommand{\themysection}{\Alph{section}}
 
In this Appendix I will give a brief review of two-component 
spinor notation \cite{mybook}.  Two-component spinors provide the
most natural spinor representations of the Lorentz group in theories
with chiral fermions, such as the standard model or supersymmetry.  The
notation exploits the fact that spinor representations of the Lorentz
group are actually two-dimensional representations of its
universal covering group, SL(2,$C$).

To begin, let us define $M$ to be a two-by-two matrix of determinant
one: $M \in$ SL(2,$C$).   The matrix $M$, its complex conjugate $M^*$,
its transpose inverse $(M^T)^{-1}$, and its hermitian conjugate inverse
$(M^\dagger)^{-1}$ are all representations of SL(2,$C$).  These
matrices  represent the action of the Lorentz group on two-component
Weyl spinors.

Two-component spinors with upper or lower dotted or undotted 
indices are defined to transform as follows under SL(2,$C$):
\begin{eqnarray}
\psi'_\alpha \;= \;M_\alpha{}^\beta\, \psi_\beta &
\qquad &
\bar{\psi}^\prime_\dalpha \;= \;M^*{}_\dalpha{}^\dbeta 
\,\bar{\psi}_\dbeta \nonumber \\
\psi^{\prime\alpha} \;= \;M^{-1}{}_{\beta}{}^\alpha\, \psi^\beta 
& \qquad & 
\bar{\psi}^{\prime\dalpha} \;= \;(M^*){}^{-1}
{}_{\dbeta}{}^{\dalpha}\, \bar{\psi}^\dbeta .
\end{eqnarray}
The spinors are denoted by Greek indices.  Those with dotted 
indices transform in the $(0,{1\over2})$ representation of the 
Lorentz group, while those with undotted indices transform 
in the $({1\over2},0)$ conjugate representation.

The map from SL(2,$C$) to the Lorentz group is established through
the $\sigma$-matrices,
\begin{equation}
\begin{array}{rclcrcl}
\sigma^0& = &\pmatrix{-1 & 0 \cr 0 & -1 \cr} & \qquad &
\sigma^1& = &\pmatrix{ 0 & 1 \cr 1 &  0  \cr}  \\[6mm]
\sigma^2& = &\pmatrix{ 0 & -i \cr i & 0 \cr} & \qquad &
\sigma^3& = &\pmatrix{ 1 & 0 \cr 0 & -1 \cr}\;,  
\end{array}
\end{equation}
The $\sigma$ matrices form a basis for two-by-two complex
matrices,
\begin{equation}
P\;\equiv\;(p_m \sigma^m)\;=\;\pmatrix{
 - p_0 + p_3 & p_1 - ip_2 \cr
 p_1 + ip_2 & -p_0 - p_3 \cr}\;.
\end{equation}
Any hermitian matrix may be expanded with the $p_m$ real.

From any hermitian matrix $P$, we may always obtain another 
by the following transformation,
\begin{equation}
P'\;=\;MPM^\dagger\;. \
\end{equation}
Both $P$ and $P'$ have expansions in $\sigma$,
\begin{equation}
(\sigma^m p'_m)\;=\;M\, (\sigma^m p_m) \,M^\dagger\;, \
\end{equation}
with $p_m$ and $p'_m$ real.  Since $M$ is unimodular (det $M$ = 1), the
coefficients $p_m$ and $p'_m$ are related by a Lorentz transformation:
\begin{equation}
{\rm det}(\sigma^m p'_m)\;=\;
{\rm det}(\sigma^m p_m)\;=\;
p_0^{\,\prime\, 2}\;-\;{\vec p}^{\;\prime\, 2}
\;=\;p_0^2\;-\;{\vec p}^{\;2}\;. \
\end{equation}
Vectors and tensors are distinguished from spinors by their Latin indices. 

From (A.1) and (A.5), we see that $\sigma^m$ has the following index
structure:
\begin{equation}
\sigma^m_{\alpha \dalpha}\;. \
\end{equation}
With these conventions, $\psi^\alpha \psi_\alpha$,
$\bar{\psi}_\dalpha \bar{\psi}^\dalpha$ and
$\psi^\alpha \sigma_{\alpha \dalpha}^m \partial_m
\bar{\psi}^\dalpha$ are all Lorentz scalars.

Because $M$ is unimodular, the antisymmetric tensors
$\epsilon^{\alpha \beta}$ and $\epsilon_{\alpha \beta}$
$(\epsilon_{21} = \epsilon^{12} = 1, \epsilon_{12} =
\epsilon^{21} = -1, \epsilon_{11} = \epsilon_{22} = 0)$
are invariant under Lorentz transformations,
\begin{eqnarray}
\epsilon_{\alpha \beta}& = &M_\alpha{}^\gamma M_\beta{}^\delta
\epsilon_{\gamma \delta} \nonumber \\
\epsilon^{\alpha \beta}& = &\epsilon ^{\gamma \delta} 
M_\gamma{}^{\alpha}
M_\delta{}^\beta \;.
\end{eqnarray}
This implies that spinors with upper and lower indices are related 
through the\linebreak $\epsilon$-tensor,
\begin{equation}
\psi^\alpha\;=\;\epsilon^{\alpha \beta}\, \psi_\beta,\qquad \psi_\alpha 
\;=\;\epsilon_{\alpha \beta}\, \psi^\beta\;.
\end{equation}
Note that we have defined $\epsilon_{\alpha \beta}$ and
$\epsilon^{\alpha \beta}$ such that $\epsilon_{\alpha \beta}
\epsilon^{\beta \gamma} = \delta_\alpha{}^\gamma$.  Analogous
statements hold for the $\epsilon$-tensor with dotted 
indices. 

The $\epsilon$-tensor may also be used to raise the indices 
of the $\sigma$-matrices,
\begin{equation}
\bsigma^{m \dalpha \alpha}\;=\;\epsilon^{\dalpha
\dbeta} \epsilon^{\alpha \beta}\, \sigma^m_{\beta
\dbeta}\;. 
\end{equation}
From the definition of the $\sigma$-matrices, we find
\begin{eqnarray}
(\sigma^m\bsigma^n + \sigma^n\bsigma^m)_{\alpha}{}^\beta 
& = & -\;2\, g^{mn}\delta_\alpha{}^\beta \nonumber \\
(\bsigma^m\sigma^n + \bsigma^n\sigma^m)^{\dalpha}{}_\dbeta 
& = & -\;2\, g^{mn}\delta^\dalpha{}_\dbeta \;,
\end{eqnarray}
and
\begin{eqnarray}
{\rm Tr}\, \sigma^m \bsigma^n& = & -\;2\, g^{mn} \nonumber \\
\sigma_{\alpha \dalpha}^{m}\bsigma_m{}^{\dbeta\beta} & = & -\;2\,
\delta_\alpha{}^\beta \delta_\dalpha{}^\dbeta \;,
\end{eqnarray}
where $g_{mn} = {\rm diag}(-1,1,1,1)$.
These relations may be used to convert a vector to a bispinor 
and vice versa:
\begin{equation}
v_{\alpha \dalpha}\;=\;\sigma_{\alpha 
\dalpha}^m\,
v_m\;, \qquad v^m = -\;{1 \over 2}\;\bsigma^{m
\dalpha\alpha}\,v_{\alpha \dalpha}\;.
\end{equation}

The generators of the Lorentz group in the spinor 
representation are given by
\begin{eqnarray}
\sigma^{nm}{}_{\alpha}{}^{\beta}& =& {1 \over 4}\, (\sigma 
_{\alpha
\dalpha}^{n} \bsigma^{m \dalpha \beta} -
\sigma_{\alpha \dalpha}^{m}\bsigma^{n 
\dalpha\beta}) \nonumber \\
\bar{\sigma}^{nm \dalpha}{}_\dbeta& =& {1\over 4}\,
(\bsigma^{n \dalpha \alpha} \sigma_{\alpha
\dbeta}^m - \bsigma^{m \dalpha 
\alpha}\sigma_{\alpha
\dbeta}^n)\;.
\end{eqnarray}
Other useful relations involving the $\sigma$-matrices are
\begin{eqnarray}
\bar{\sigma}^a \sigma^b \bar{\sigma}^c\;-\;\bar{\sigma}^c
\sigma^b \bar{\sigma}^a& =& - 2i\, \epsilon^{abcd}
\bar{\sigma}_d \nonumber \\
\sigma^a \bar{\sigma}^b \sigma^c\;-\;\sigma^c \bar{\sigma}^b
\sigma^a& =& 2i \epsilon^{abcd} \sigma_d \;,
\end{eqnarray}
where $\epsilon_{0123} = - 1$, as well as
\begin{eqnarray}
\sigma^a \bar{\sigma}^b \sigma^c\;+\;\sigma^c \bar{\sigma}^b
\sigma^a& =& 2(g^{ac} \sigma^b\;-\;g^{bc} \sigma^a\;-\;g^{ab} 
\sigma^c)
\nonumber \\
\bar{\sigma}^a \sigma^b \bar{\sigma}^c\;+\;\bar{\sigma}^c 
\sigma^b
\bar{\sigma}^a& =& 2(g^{ac} \bar{\sigma}^b\;-\;g^{bc} 
\bar{\sigma}^a\;-
\;g^{ab} \bar{\sigma}^c)
\end{eqnarray}
and
\begin{eqnarray}
\sigma_{\alpha \dalpha}^n\sigma_{\beta 
\dbeta}^m\;-\
\sigma_{\alpha \dalpha}^m \sigma_{\beta \dbeta}^n 
& = &
2[(\sigma^{nm}\epsilon)_{\alpha \beta}
\epsilon_{\dalpha\dbeta}\;+\;(\epsilon
\bar{\sigma}^{nm})_{\dalpha \dbeta}
\epsilon_{\alpha\beta}] \nonumber \\
\sigma_{\alpha \dalpha}^n \sigma_{\beta \dbeta}^m\;+\;
\sigma_{\alpha \dalpha}^m \sigma_{\beta \dbeta}^n
& =&
-\;g^{nm} \epsilon_{\alpha \beta}
\epsilon_{\dalpha\dbeta}\;+\;4(\sigma^{\ell
n}\epsilon)_{\alpha \beta} (\epsilon \bar{\sigma}^{\ell
m})_{\dalpha\dbeta}\;.
\end{eqnarray}

Equation (A.11) makes it easy to relate two-component to
four-component spinors.  This is done through the following
realization of the Dirac $\gamma$-matrices:
\begin{equation}
\gamma^m\;=\;\pmatrix{
0 & \sigma^m \cr
\bar{\sigma}^m & 0 \cr}\;.
\end{equation}
We call this the Weyl basis.  In this basis, Dirac 
spinors contain two Weyl spinors,
\begin{equation}
\Psi_D\;=\;\pmatrix{
\chi_\alpha \crr
\bar{\psi}^\dalpha \crr}\;,
\end{equation}
while Majorana spinors contain only one:
\begin{equation}
\Psi_M\;=\;\pmatrix{
\chi_\alpha \crr
\bar{\chi}^\dalpha \crr}\;.
\end{equation}

Throughout these lectures we will use the following spinor
summation convention,
\begin{eqnarray}
&\psi \chi\;=\;\psi^\alpha \chi_\alpha\;=\;- \psi_\alpha 
\chi^\alpha\;=\
\chi^\alpha \psi_\alpha\;=\;\chi \psi & \nonumber \\
& \bar{\psi} \bar{\chi}\;=\;
\bar{\psi}_{\dalpha}\bar{\chi}^{\dalpha}\;=\;
- \bar{\psi}^{\dalpha}\bar{\chi}_{\dalpha}\;=\;
\bar{\chi}_{\dalpha}\bar{\psi}^{\dalpha}\;=\;
\bar{\chi} \bar{\psi}\;.&
\end{eqnarray}
Here we have assumed, as always, that spinors anticommute.  
The definition of $\bar{\psi} \bar{\chi}$ is chosen in such a way 
that
\begin{equation}
(\chi \psi)^\dagger\;=\;(\chi^\alpha \psi_\alpha)^\dagger = 
\bar{\psi}_\dalpha
\bar{\chi}^\dalpha\;=\;\bar{\chi} \bar{\psi}\;. 
\end{equation}
Note that conjugation reverses the order of the spinors.

\end{document}